\documentclass[sigconf]{acmart}

\graphicspath{{figs_cameraready/}}

\usepackage{subcaption}
\usepackage{multirow}
\usepackage{xcolor}
\usepackage{colortbl}

\usepackage{pifont}
\newcommand{\cmark}{\ding{51}}

\usepackage{mathtools}

\newcolumntype{C}[1]{>{\centering\let\newline\\\arraybackslash\hspace{0pt}}m{#1}}

\AtBeginDocument{%
  \providecommand\BibTeX{{%
    \normalfont B\kern-0.5em{\scshape i\kern-0.25em b}\kern-0.8em\TeX}}}

\copyrightyear{2024}
\acmYear{2024}
\setcopyright{rightsretained}
\acmConference[DIS '24]{Designing Interactive Systems Conference}{July 1--5, 2024}{IT University of Copenhagen, Denmark}
\acmBooktitle{Designing Interactive Systems Conference (DIS '24), July 1--5, 2024, IT University of Copenhagen, Denmark}\acmDOI{10.1145/3643834.3661594}
\acmISBN{979-8-4007-0583-0/24/07}

\begin{document}

\title{DanceGen: Supporting Choreography Ideation and Prototyping with Generative AI}

\author{Yimeng Liu}
\affiliation{
  \institution{University of California, Santa Barbara}
  \city{Santa Barbara}
  \postcode{93106}
  \country{USA}}
\email{yimengliu@cs.ucsb.edu}

\author{Misha Sra}
\affiliation{
  \institution{University of California, Santa Barbara}
  \city{Santa Barbara}
  \postcode{93106}
  \country{USA}}
\email{sra@cs.ucsb.edu}

\renewcommand{\shortauthors}{Liu and Sra}


\begin{abstract}
Choreography creation requires high proficiency in artistic and technical skills. Choreographers typically go through four stages to create a dance piece: preparation, studio, performance, and reflection. This process is often individualized, complicated, and challenging due to multiple constraints at each stage. To assist choreographers, most prior work has focused on designing digital tools to support the last three stages of the choreography process, with the preparation stage being the least explored. To address this research gap, we introduce an AI-based approach to assist the preparation stage by supporting ideation, creating choreographic prototypes, and documenting creative attempts and outcomes. We address the limitations of existing AI-based motion generation methods for ideation by allowing generated sequences to be edited and modified in an interactive web interface. This capability is motivated by insights from a formative study we conducted with seven choreographers. We evaluated our system's functionality, benefits, and limitations with six expert choreographers. Results highlight the usability of our system, with users reporting increased efficiency, expanded creative possibilities, and an enhanced iterative process. We also identified areas for improvement, such as the relationship between user intent and AI outcome, intuitive and flexible user interaction design, and integration with existing physical choreography prototyping workflows. By reflecting on the evaluation results, we present three insights that aim to inform the development of future AI systems that can empower choreographers. 
\end{abstract}



\maketitle


\section{Introduction}
The creation of dance choreography is a complex, multifaceted, and embodied process that demands a high level of artistic and technical skill. It requires a comprehension of music and movement and the ability to create and communicate a clear artistic vision~\cite{kogan2002careers}. Choreographers must consider the physical abilities and limitations of their dancers, the emotional tone of the piece, and the overall aesthetic of the production to ensure the outcome is both visually appealing and emotionally resonant~\cite{calvert1989composition, kirsh2009choreographic, fdili2017seeing}. 

Although the choreography process can vary across individuals, prior research~\cite{ciolfi2016choreographers} has summarized the typical choreography process, which consists of four main stages: \textit{preparation} of materials, collaboration with dancers in a \textit{studio}, dance \textit{performance}, and \textit{reflection}. This process is non-linear and occurs recursively within each stage and between different stages. During choreography, common challenges include staying creative and avoiding clich\'{e}s, dealing with stressful time constraints on creating routines quickly, creating for dancers with different physical abilities and ranges of motion, and receiving and integrating feedback from collaborators, such as directors and production designers. 
To help address some of these challenges, previous work has designed systems using both non-AI and AI approaches. Non-AI choreography support systems, such as~\cite{ladenheim2020live, molina2017delay, oulasvirta2013information}, have been shown to be effective in helping the analysis of dance movements and extending the body expressivity of dancers. Most of these systems have been designed to support the studio, performance, and reflection stages in the choreography process, but few aid the preparation stage that focuses on ideation and intense brainstorming. AI choreography support methods, on the other hand, have demonstrated their benefits in assisting the preparation stage by allowing the generation of choreography materials. For instance, TM2D~\cite{gong2023tm2d} generates dance sequences from text and music, and Everybody Dance Now~\cite{chan2019everybody} generates new dance videos by animating full-body human images with user-selected dance videos. These AI models are responsive and can generate dance sequences on demand, providing a valuable resource for building choreography materials. Furthermore, generative AI algorithms, such as diffusion models, can generate unusual dance movements. Such artifacts can potentially spark creativity, as investigated by prior research~\cite{carlson2015moment, rothenberg2019error}. However, most of these AI methods fail to allow the iterative editing of generated results, a crucial need in the preparation stage of the choreography process. 

\begin{figure*}[!ht]
    \centering
    \includegraphics[width=0.88\textwidth]{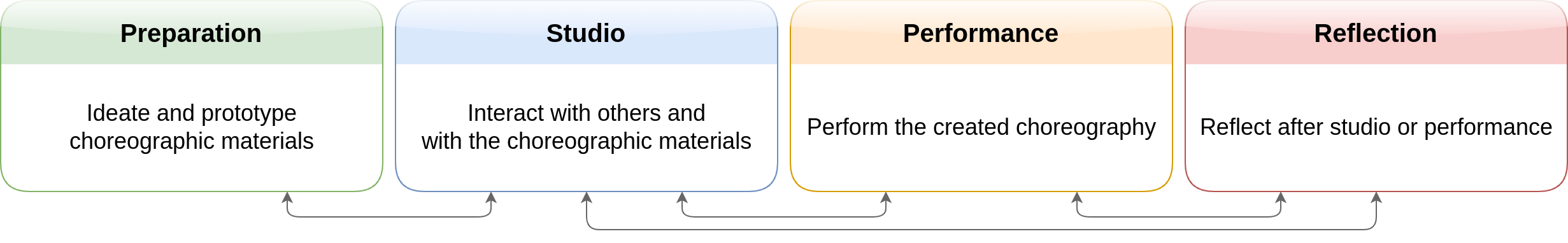}
    \caption{Choreography creation process, as outlined by prior research~\cite{ciolfi2016choreographers}, consisting of four stages: \textit{preparation}, \textit{studio}, \textit{performance}, and \textit{reflection}. These stages are interconnected and occur recursively, as indicated by the arrows.}
    \Description{This flowchart illustrates the choreography creation process, consisting of four stages represented by four horizontally aligned blocks. From left to right, these stages are preparation, studio, performance, and reflection. Double-ended arrows indicate the flow between stages, encompassing transitions from preparation to studio, studio to performance, performance to reflection, and studio to reflection.}
    \label{fig:choreography_process}
\end{figure*}

Given the limited prior work on assisting the preparation stage and known AI benefits to support choreography ideation, we aim to fill this research gap by harnessing generative AI to support the choreography preparation stage. To better understand the design specifications for a digital choreography-support system, we conducted a formative study with seven experienced participants to formulate our design goals. The study identified challenges, needs, and opportunities to design a digital system to help choreography preparation. According to our findings, an ideal system needs to support efficient ideation that offers choreographers a wide range of choreography materials to spark novel ideas. It also needs to aid digital prototyping and rapid iteration for early testing of choreographic ideas and reduce the demand for physical prototyping, which is otherwise time-consuming and expensive. Lastly, it needs to allow choreographers to document ideas and dance sequences for reflection and reuse, as well as share the documentation with collaborators to communicate their creative vision, get feedback, and initiate collaborations. 

Based on the design goals, we present DanceGen, a system to support choreography preparation. We adopt a generative AI algorithm to aid ideation, which allows choreographers to generate dance sequences based on natural language descriptions. Our AI model is a fine-tuned version of the human motion generation model, MDM~\cite{tevet2023human}, with modifications to support dance generation. We also allow iterative editing of AI-generated outcomes through an interactive web-based user interface. Lastly, we enable the documentation of intermediate steps and polished results, including the commands to generate and edit dance motion and final dance sequences as videos and 3D animated motions. We evaluated the usability of our system by conducting a user study with six expert choreographers. We share participant feedback on system functionality and their insights into how the system can be used to support creative exploration in the choreography preparation stage. We further analyze the results and discuss lessons learned from the design and evaluation of the system.

Our main contributions are as follows: (1) results from a formative study that helps us understand choreographer needs and challenges to formulate our design goals; (2) a generative AI-supported system with a web user interface that enables the generation, iterative prototyping, and 2D and 3D documentation of dance sequences; (3) findings from a system usability evaluation, offering insights into the functionality of our system and its potential use to assist the choreography preparation stage; (4) a discussion based on analyzing user evaluation results, which may inform the development of future AI choreography-support systems.

\section{Background: Choreography Creation Process} \label{sec:background}
Choreography embodies a highly creative process, which tends to be unique across different choreographers and evolves over time for an individual choreographer. Previous literature~\cite{calvert1989composition, alaoui2014choreography, ciolfi2016choreographers} indicates the absence of a standardized procedure or theoretical framework in the choreography creation domain. However, according to Ciolfi et al.~\cite{ciolfi2016choreographers}, the choreography process follows a general design process derived from extensive interviews with professional choreographers. Specifically, it comprises of four stages: \textit{preparation}, \textit{studio}, \textit{performance}, and \textit{reflection}, as depicted in Figure~\ref{fig:choreography_process}. The process typically flows iteratively between pairs of these stages, including transitions between preparation and studio, studio and performance, performance and reflection, and studio and reflection.

At the beginning of the choreography process, choreographers receive a brief, such as a specific theme for a particular purpose. They are also informed of any constraints, including the style, number of dancers, stage space, music, timeline, and any required props. Armed with the foundational information, they start by generating a set of choreographic ideas. To express, visualize, and record ideas, choreographers utilize various approaches, such as textual descriptions, videos, physical demonstrations, sketches, diagrams, and formal notation~\cite{zhou2021dance}. 
When choreographers face challenges in expressing their ideas satisfactorily, they often turn to online videos or seek advice from their peers. 
Communicating choreographic ideas is crucial for collaboration with fellow choreographers or dancers. This facilitates the transmission of choreographic ideas from the preparation stage to the studio and performance stages or for future reflection. 
In the choreography creation process, choreographic ideas and dance movements undergo refinement and evolve iteratively. Choreographers incorporate new material or feedback, continually assessing the motions until the piece is satisfactory.

\section{Related Work} \label{sec:related_work}
In this section, we summarize existing choreography-support systems to help position our work. We classify these systems into non-AI and AI-based categories and outline them in Table~\ref{tab:rw_comparison} for comparison. 

\begin{table*}[!ht]
    \centering
    \caption{Comparison of choreography-support systems in prior work. Our work focuses on the preparation stage and details the specific aspects our system supports in this stage for comparison.}
    \scalebox{0.88} {
    \begin{tabular}{cp{0.12\textwidth}lcC{0.1\textwidth}C{0.1\textwidth}C{0.1\textwidth}}
    \toprule
        \multirow{2.5}{*}{Type} & \multirow{2.5}{*}{Paper} & \multirow{2.5}{*}{Description} & \multicolumn{4}{c}{Choreography Process}\\
         \cmidrule{4-7}
         &  &  & Preparation & Studio & Performance & Reflection\\
        \midrule
        \multirow{15.5}{*}{Non-AI} &\cite{karpashevich2018reinterpreting, johnston2015conceptualising, jochum2019tonight, gemeinboeck2017movement, ladenheim2020live, allen2022wearable} & \begin{tabular}{@{}p{0.22\textwidth}@{}} Accompany dancers with novel technology \end{tabular} & \cellcolor{gray!15} &  \cmark &  \cmark & \cellcolor{gray!15} \\\cmidrule{2-7}
         &\cite{janauskaite2019establishing, eriksson2019dancing, raheb2018choreomorphy} & \begin{tabular}{@{}p{0.22\textwidth}@{}} Create novel visualization for dance performance \end{tabular} & \cellcolor{gray!15} &  \cmark &  \cmark & \cellcolor{gray!15} \\\cmidrule{2-7}
         &\cite{oulasvirta2013information, velloso2013motionma, fdili2015experts} & \begin{tabular}{@{}p{0.22\textwidth}@{}} Analyze abstract physical movements \end{tabular} & \cellcolor{gray!15} & \cellcolor{gray!15} & \cellcolor{gray!15} &  \cmark\\\cmidrule{2-7}
         &\cite{singh2011choreographer, carlson2015moment} & \begin{tabular}{@{}p{0.22\textwidth}@{}} Annotate and sketch physical movements \end{tabular} & \cellcolor{gray!15} &  \cmark & \cellcolor{gray!15} &  \cmark\\\cmidrule{2-7}
         & \begin{tabular}{@{}p{0.12\textwidth}@{}}\cite{calvert1989composition, ciolfi2018knotation, raheb2018choreomorphy, molina2017delay, carlson2019shifting, franccoise2022co} \end{tabular} & \begin{tabular}{@{}p{0.22\textwidth}@{}} Spark novel ideas and support documentation \end{tabular} & \begin{tabular}{@{}p{0.29\textwidth}@{}}  \textit{Ideation}: Dance motion display \\ \textit{Prototyping}: Physical \\ \textit{Documentation}: Video, sketching \end{tabular} &  \cmark & \cellcolor{gray!15} & \cellcolor{gray!15} \\\cmidrule{2-7}
         &\cite{singh2011choreographer, carlson2015sketching, zhou2023here, gemeinboeck2017movement} & \begin{tabular}{@{}p{0.22\textwidth}@{}} Collaborate with dancers and choreographers \end{tabular} & \cellcolor{gray!15} &  \cmark &  \cmark & \cellcolor{gray!15} \\
        \midrule
        \multirow{8}{*}{AI} &\cite{chuang2022music2dance, zhang2022music, alexanderson2023listen, ye2020choreonet, tseng2023edge} & \begin{tabular}{@{}p{0.22\textwidth}@{}} Generate dance based on music \end{tabular} & \multirow{4}{*}{\begin{tabular}{@{}p{0.29\textwidth}@{}} \textit{Ideation}: Dance motion generation \\ \textit{Prototyping}: N/A \\ \textit{Documentation}: Video, 3D \end{tabular}} & \cellcolor{gray!15} & \cellcolor{gray!15} & \cellcolor{gray!15} \\
        \cmidrule{2-3}\cmidrule{5-7}
         &\cite{gong2023tm2d} & \begin{tabular}{@{}p{0.22\textwidth}@{}} Generate dance based on text \end{tabular} &  & \cellcolor{gray!15} & \cellcolor{gray!15} & \cellcolor{gray!15} \\
         \cmidrule{2-3}\cmidrule{5-7}
         &\cite{chan2019everybody} & \begin{tabular}{@{}p{0.22\textwidth}@{}} Generate dance based on video \end{tabular} &  & \cellcolor{gray!15} & \cellcolor{gray!15} & \cellcolor{gray!15} \\\cmidrule{2-7}
         & DanceGen (ours) & \begin{tabular}{@{}p{0.22\textwidth}@{}} Generate and refine dance based on text and video \end{tabular} & \begin{tabular}{@{}p{0.29\textwidth}@{}}  \textit{Ideation}: Dance motion \textbf{generation} and \textbf{modification} \\ \textit{Prototyping}: \textbf{Digital}, \textbf{editable}, \textbf{iterative} \\ \textit{Documentation}: \textbf{Video}, \textbf{3D}, \textbf{text} \end{tabular} & \cellcolor{gray!15} & \cellcolor{gray!15} & \cellcolor{gray!15} \\
    \bottomrule
    \end{tabular}
    }
    \Description{This table categorizes existing choreography-support systems into non-AI-based and AI-based, listing corresponding papers along with summaries. The table includes checkmarks indicating which choreography stages the papers support. As observed, most non-AI-based systems have been designed to support the studio, performance, and reflection stages, while AI-based systems have demonstrated effectiveness in supporting the preparation stage.}
    \label{tab:rw_comparison}
\end{table*}

\subsection{Non-AI Choreography-support Systems}
The recent survey paper by Zhou et al.~\cite{zhou2021dance} provides a comprehensive summary of choreography-support systems developed by HCI researchers over the past two decades. According to this paper, non-AI choreography-support systems can be categorized into three main groups: those that enhance body movements by computing to accompany dancers and create novel visual effects, those for annotating and analyzing dance sequences to understand and modify abstract movements, and those supporting the overall choreography process.
In the first category, past research has explored the design of robotic or algorithmic agents that accompany human dancers~\cite{karpashevich2018reinterpreting, johnston2015conceptualising, jochum2019tonight, gemeinboeck2017movement, ladenheim2020live, allen2022wearable} and the creation of innovative visualizations for dance performances~\cite{janauskaite2019establishing, eriksson2019dancing, raheb2018choreomorphy}. 
In the second category, researchers have utilized motion capture technology to analyze dance movement quality and emotional expressiveness~\cite{oulasvirta2013information, velloso2013motionma, fdili2015experts}. Additionally, tools have been developed to annotate and sketch movements, supporting the initiation or modification of dance sequences~\cite{singh2011choreographer, carlson2015moment}. 
In the last category, previous work has focused on designing systems and tools to inspire and document new ideas~\cite{calvert1989composition, ciolfi2018knotation, raheb2018choreomorphy, molina2017delay, carlson2019shifting, franccoise2022co} and facilitate collaboration among dance production teams~\cite{singh2011choreographer, carlson2015sketching, zhou2023here, gemeinboeck2017movement}.

While these systems have demonstrated effectiveness in augmenting choreography creation from multiple aspects, they mostly focused on supporting the studio, performance, and reflection stages of the choreographic process, leaving the preparation stage largely underexplored. The preparation stage, emphasizing ideation and prototyping before dance-making in a studio, demands imagination, creativity, experience, and knowledge to generate innovative ideas and craft materials for the latter three stages to build upon. Meeting all these demands is non-trivial for choreographers, requiring wise time and effort management for success. Previous research has underscored the need for newer technologies to augment creativity~\cite{ciolfi2016choreographers, zhou2021dance} and reduce physical strain~\cite{calvert1989composition} in this stage. Our work aims to address these challenges by designing a system that assists ideation, digital prototyping, and documentation targeted at creative activities performed during the preparation stage. 

\subsection{AI Choreography-support Methods} \label{sec:rw_ai}
AI choreography-support methods can be advantageous in facilitating the choreography preparation stage. Generative AI algorithms designed for dance and motion generation have demonstrated effectiveness in creating complex and realistic motion sequences, employing various backbone structures, such as graph convolutional networks (GCNs)~\cite{ferreira2021learning, ren2020self, yan2019convolutional}, generative adversarial networks (GANs)~\cite{lee2019dancing, sun2020deepdance}, transformers~\cite{Li_2021_ICCV, perez2021transflower, liu2021motion, Siyao_2022_CVPR, li2020learning}, and diffusion models~\cite{li2023finedance, qi2023diffdance, zhou2023ude}. Among these structures, diffusion models stand out for their ability to generate motions with variety.

Existing generative AI approaches have explored utilizing multiple input modalities for dance generation. One prevalent input option is music, where AI models have converted music features such as style, rhythm, and beat into dance motions~\cite{chuang2022music2dance, zhang2022music, alexanderson2023listen, ye2020choreonet, tseng2023edge}. While music is commonly used as an input modality for choreography creation, it is not a mandatory requirement, and choreographers may even purposefully choose to create dance movements independently of music~\cite{stevens2009moving, mason2012music}. In some cases, choreographers find it beneficial to withhold music to avoid dependence on musical elements when creating new movements. Alternatively, choreographers have employed input types, such as gestures, physical demonstration, and vocalization, to create dance~\cite{kirsh2009choreographic}. Text serves as another useful modality to convey choreographic ideas, and it has been effectively used for dance generation by generative AI methods~\cite{gong2023tm2d}. Learning the mapping between language and dance has been shown to be desirable due to its zero-shot capability~\cite{jiang2023motiongpt} and using natural language to communicate ideas aligns with the existing practices of choreographers~\cite{ciolfi2016choreographers}. 
Text input is also easily editable, facilitating nuanced control to generate varied dance sequences. Additionally, prior research has employed dance videos to animate human full-body images for dance generation~\cite{chan2019everybody}. While less flexible to edit than text, videos have been frequently used to document, communicate, and share choreographic ideas~\cite{ciolfi2016choreographers, zhou2021dance}. 

The AI choreography-support methods discussed here share the common advantage of quickly generating diverse dance motions through multiple input modalities, making them ideal for brainstorming and ideation. This process converts abstract descriptions, such as text and music, into detailed representations --- dance sequences. This capability can benefit the preparation stage by offering a way to generate choreographic materials rapidly. In this work, we present a system that leverages generative AI and allows the use of text and visual data as input for dance generation. Specifically, our system is built on a diffusion-based text-to-motion generative neural network, MDM~\cite{tevet2023human}, known for its superior performance and ability to generate a wide range of motions. However, MDM and many existing AI choreography-support methods only allow one-time automatic motion generation and do not let users edit or customize the generated outcomes. We address this limitation by enabling the editing of generated results, which facilitates iterative prototyping and aligns with the typical choreography creation process introduced in Section~\ref{sec:background}.

\section{Formative Study} \label{sec:formative_study}
To determine the design specifications for a system that can support the choreography preparation stage, we conducted a formative study. Our goal was to gain insights into the needs and challenges of choreographers and identify opportunities to design a new system. We held semi-structured interviews with individuals experienced in dance-making and extracted key findings to establish design goals that helped guide the development of our system.

\subsection{Participants} 
To capture a broad range of perspectives, participant recruitment encompassed a variety of choreography backgrounds and purposes, enriching the knowledge base for our system's development. We recruited seven participants (Female: 6, Male: 1) with choreography experience and assigned them unique IDs from FS-P01 to FS-P07. Their dance choreography expertise included different styles and genres, such as ballet, contemporary, street, ballroom, and classical dance. They had choreography experience from three to fifteen years, from creating dances for teaching and assignments to performances and personal projects. 

\subsection{Procedure}
The study has been approved by the local Institutional Review Board (IRB) under protocol \# 17-23-0083. Participant interactions occurred in person or over Zoom, depending on their availability and preference. They provided informed consent and were briefed on the study's purpose. Subsequently, a set of questions was posed, covering the participant's own creative process, the choreographic tools and methods they employed, and the challenges they encountered. The complete interview questions are available in Appendix~\ref{sec:formative_study_interview}.

\subsection{Data Collection and Analysis}
We documented participant responses to each question through hand-written notes. The collected data was subsequently analyzed using thematic analysis~\cite{braun2006using, braun2012thematic}, which involved coding and categorizing common themes in the interview data. This process aimed to extract design implications to inform the design of our system.

\subsection{Findings}
The data analysis revealed three main themes, covering the challenges and needs in our participants' current choreography process. These themes are grounded in the specific choreography experiences of the participants. 

\subsubsection{Sparking Inspiration for Creativity.}
Maintaining a steady flow of creative inspiration is a persistent challenge, even for choreographers with years of experience. To keep their creative fires burning, choreographers need to continuously seek new information and experiences. Common methods can be categorized as either active or passive. Active approaches require conscious effort, such as searching online resources, consulting fellow artists, learning new dance styles (FS-P02), exploring different music (FS-P01), and engaging in improvisation (FS-P03). Passive inspiration, on the other hand, can strike spontaneously through everyday experiences or exposure to various art forms in addition to dance, like visiting museums or attending concerts (FS-P05).

\subsubsection{Managing Time and Effort for Creative Success.}
Choreography creation poses a distinct challenge in managing time and effort, whether working individually or collaboratively on projects ranging from short TikTok videos to extensive stage productions. Regardless of the desired outcome, developing fresh choreographic ideas, refining a dance piece, and producing the final outcome demand significant mental and physical investment. Effectively managing time and effort across a multitude of tasks, such as brainstorming, communication with other choreographers and dancers, and physical rehearsal (FS-P06), becomes crucial for success. These tasks often require multiple iterations (FS-P06), further amplifying the need for wise time management within the allocated timeframe.

\subsubsection{Capturing Movement for Reflection and Collaboration.}
Accurate and timely documentation of dance movements is crucial in the choreography process, as choreographers frequently use their prior choreography as a springboard to generate new ideas. Specifically, choreographers rely on documentation to revisit their creations, analyze the effectiveness of created pieces, and refine their artistic vision (FS-P04). Beyond self-reflection, documentation bridges collaboration (FS-P01). It allows choreographers to seek advice and insights from fellow choreographers, test and refine pieces with dancers, and communicate their vision to collaborators, ensuring everyone works towards the same artistic goals.

\begin{figure}[!ht]
    \centering
    \includegraphics[width=0.47\textwidth]{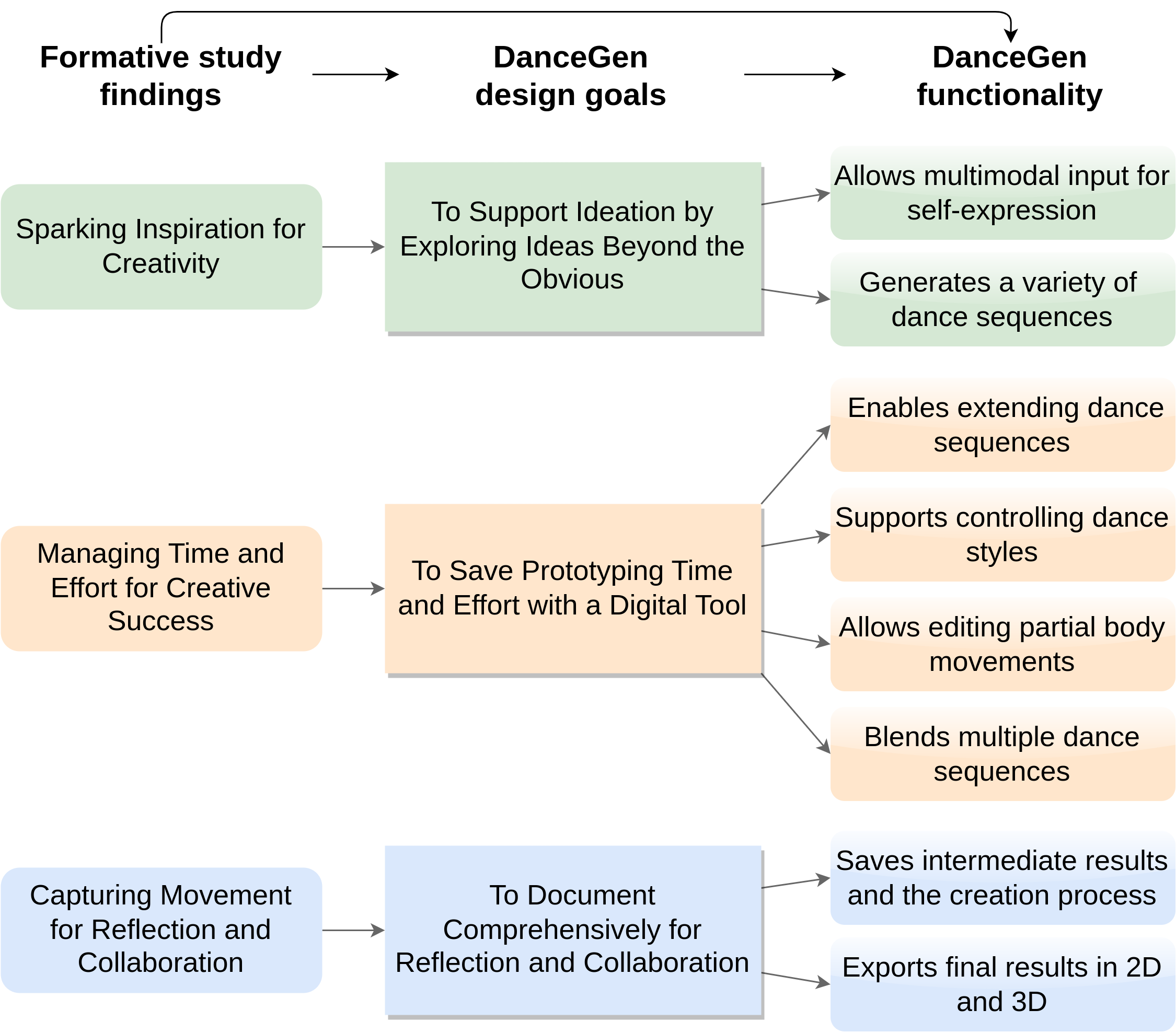}
    \caption{DanceGen design process. The formative study findings and our design goals are presented in Section~\ref{sec:formative_study}. These design goals subsequently guide the development of the system's functionality. Additionally, feedback from the formative study, including participants' choreographic methods and challenges, motivates certain system features. Detailed information on the DanceGen system and its functionality is introduced in Section~\ref{sec:dancegen_system}.}
    \Description{The design process of DanceGen is illustrated as a flowchart containing three columns: formative study findings, DanceGen design goals, and DanceGen functionality. Proceeding from left to right, the formative study findings informed the DanceGen design goals, and both the findings and design goals played a role in shaping the DanceGen functionality.
    In each column, three clusters of flowcharts, distinguished by green, orange, and blue colors, depict the design process of DanceGen, stemming from the three primary findings of the formative study. Each formative study finding, outlined in the first column, led to the design goals, introduced in the second column and, ultimately, the functionality outlined in the last column.}
    \label{fig:designing_dancegen}
\end{figure}

\subsection{Design Goals} \label{sec:design_goals}
Based on the formative study findings, we outline our design goals for a new choreography-support system. These goals aim to address the needs and challenges identified by the participants, specifically focusing on assisting the \textit{preparation stage} of the four-stage choreography creation process. We aim to support ideation, prototyping, and documentation to equip choreographers with materials and ideas for the remaining three stages. Figure~\ref{fig:designing_dancegen} summarizes the study findings and our design goals. 
Our intent is not to replace the creative efforts of choreographers but to offer a tool that aids the existing choreography process, enhancing the efficiency and creative possibilities available to choreographers. 

\subsubsection{To Support Ideation by Exploring Ideas Beyond the Obvious.}
Maintaining a steady flow of creative inspiration is a constant challenge for choreographers. To assist with idea generation, the system can provide diverse ways for choreographers to express their needs, such as text descriptions and video uploads of dance sequences. This can allow them to explore a variety of movement styles and choreographic ideas based on their input. Furthermore, the system can generate unexpected and surprising content, pushing them out of their comfort zones and sparking new creative possibilities. Exploring diverse options and encountering unexpected suggestions can be a powerful tool for overcoming creative plateaus.

\subsubsection{To Save Prototyping Time and Effort with a Digital Tool.}
Creating a new dance piece often involves time-consuming and repetitive physical rehearsals. Our goal is to reduce the need for physical prototyping by enabling rapid digital prototyping during the preparation stage. The system can allow choreographers to quickly test and refine choreographic ideas and materials, saving them valuable time and effort. Prioritizing speed and responsiveness, an intuitive user interface is desired to ensure efficient workflows and reduce learning curves, optimizing the use of limited time.

\subsubsection{To Document Comprehensively for Reflection and Collaboration.}
Accurate and detailed documentation is essential for both self-reflection and collaboration during choreography creation. To this end, the system needs to offer proper documentation methods to capture movement and creative attempts. We aim to move beyond traditional 2D documentation limitations, like video occlusions (FS-P01), by leveraging the multi-view advantages of 3D documentation~\cite{anjos2018three}. While 3D methods offer high accuracy, they often lack easy editability. To address this, the system can integrate flexible documentation techniques, such as textual descriptions, alongside visual elements, allowing for both comprehensive capture and easy refinement of the documented ideas.

\begin{figure*}[!ht]
    \centering
    \includegraphics[width=0.98\textwidth]{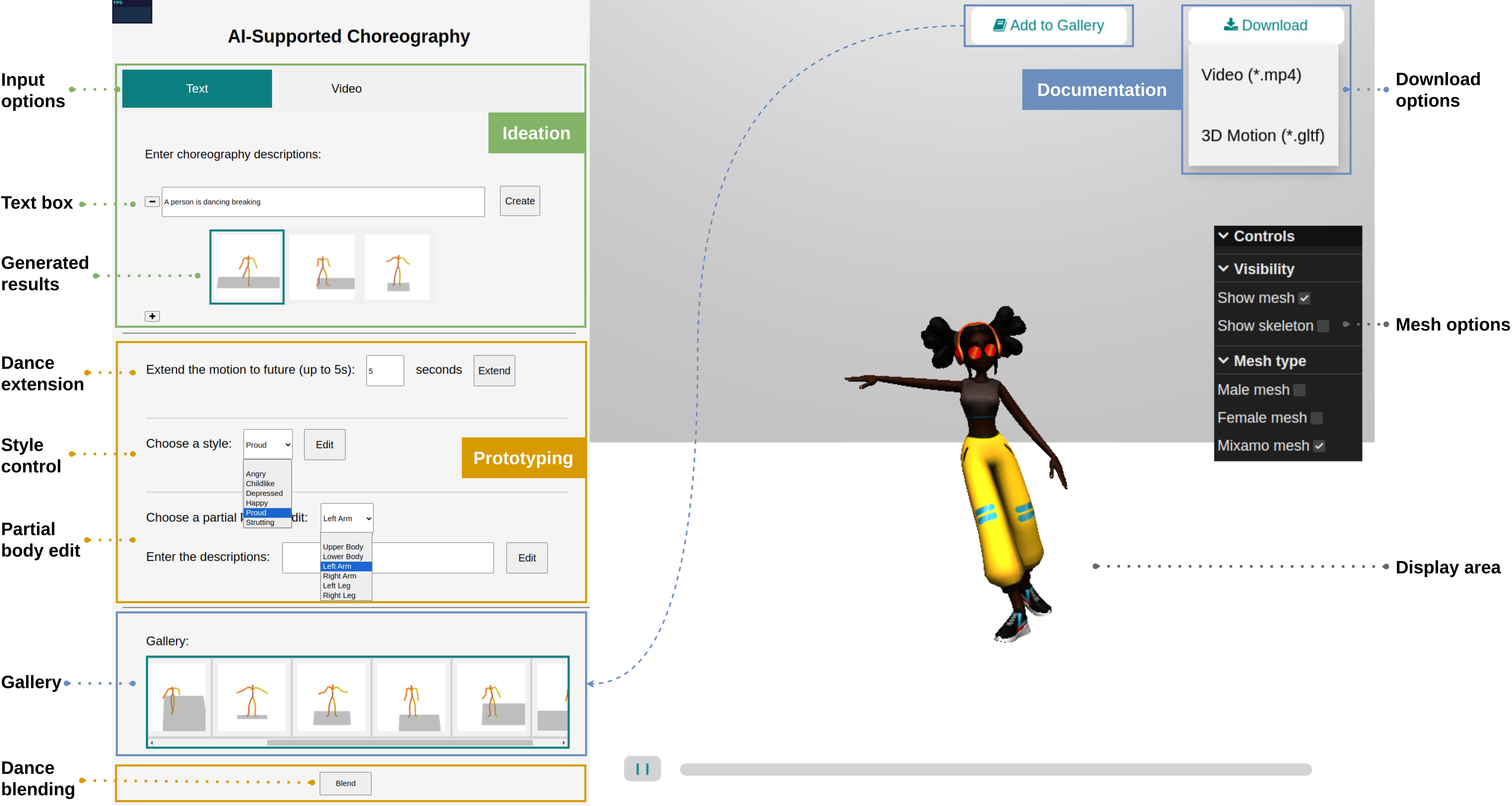}
    \caption{DanceGen user interface and functionality. On the left, users can generate dance sequences for ideation based on text descriptions. They can further edit dance sequences using the editing options for prototyping. Specifically, dance extension is facilitated by specifying the desired extension length, up to 5 seconds. Style control is achieved through a drop-down menu for users to select different movement styles. Partial body movement editing is allowed by choosing a body part and providing corresponding descriptions in the text box below. After creating dance sequences and adding them to the \textit{Gallery}, users can select dance sequences to blend them. On the right, users can view and interact with the generated dance sequences, represented by a digital avatar. The visibility of the avatar's mesh and skeleton can be adjusted using checkboxes on the far right. The current prototype offers three types of avatar meshes. Lastly, users can download the generated dance sequences in both 2D and 3D formats.}
    \Description{The DanceGen user interface depicts the editing options on the left. From top to bottom, the editing options include the dance extension option with a text box taking the extension length in seconds; the style control options in a dropdown menu containing six styles: angry, childlike, depressed, happy, proud, and strutting; and the partial body editing options in a dropdown menu containing body parts: upper body, lower body, left arm, right arm, left leg, and right leg. Below these editing options is a Gallery containing user-created dance sequences shown as thumbnails. At the bottom is a Blend button to blend two dance sequences picked from the Gallery.
    The user interface displays the system-generated dance sequences on the right. A humanoid avatar with yellow pants and red glasses is animated to demonstrate the dance movements.
    The system's features for ideation, prototyping, and documentation are grouped by bounded rectangles colored green, orange, and blue, which correspond to the colors of clusters shown in Figure~\ref{fig:designing_dancegen}.}
    \label{fig:ui_w_editoptions}
\end{figure*}

\section{DanceGen System} \label{sec:dancegen_system}
As introduced in the previous section, Figure~\ref{fig:designing_dancegen} provides an overview of the DanceGen design process. The system functionality is shaped by the design goals, and the specific features are motivated by the formative study findings. In this section, we present the system user interface in Section~\ref{sec:user_interface} and provide a detailed introduction of the functionality geared towards the design goals in Section~\ref{sec:functionality}. The technical implementation of the system functionality is briefly covered in Section~\ref{sec:functionality}, and the full technical details of the user interface and generative AI model can be found in Appendix~\ref{sec:technical_details}.

\subsection{User Interface} \label{sec:user_interface}
\subsubsection{Overview.}
Figure~\ref{fig:ui_w_editoptions} illustrates the user interface, highlighting the main functionality --- ideation, prototyping, and documentation, corresponding to the design goals. 
The input options are organized into two tabs: \textit{Text} and \textit{Video}. Selecting the text tab offers users a text input box to describe dance sequences for generation. The editing options below allow adjustments to dance sequences, including the expansion of dance sequences, alteration of dance styles, editing of partial body movements, and blending of dance sequences. Opting for the video tab replaces the text input box with a file upload button to import dance videos while the editing options remain the same.
The display area on the right shows dance sequences represented by an animated digital mesh. Currently, users can choose between SMPL male and female~\cite{SMPL_2015} meshes or a Mixamo mesh~\cite{mixamo}, which can be easily expanded to include others. The interactive meshes allow users to rotate, scale, and move them. Visibility adjustments for the mesh and skeleton can be done using the corresponding checkboxes. The \textit{Add to Gallery} button includes the user's creation into the \textit{Gallery} for review and reuse. The \textit{Download} button enables users to export the animated mesh in \textit{.gltf} format for 3D use and \textit{.mp4} format for videos.

\subsubsection{User Interface Platform.} \label{sec:ui_platform}
The system currently has a web-based user interface. However, our adaptable backend allows for potential extensions to support different frontend interfaces on various platforms, including AR/VR/MR. This adaptability opens up possibilities for immersive and interactive dance sequence generation, display, and editing. The decision to start with a web interface was based on several considerations: 
(1) Web interfaces are widely used and familiar to a broad audience, including our target users --- choreographers. 
(2) The mouse and keyboard support in web interfaces facilitates easy and rapid interactive operations, such as typing, clicking, and dragging, which are necessary to generate, edit, and view dance sequences. These operations can be more challenging in extended reality (XR) interfaces, where mid-air interactions can lead to fatigue~\cite{evangelista2021xrgonomics}, and typing one letter at a time is slow and inefficient. While speech-based input is a valid alternative, it introduces its own challenges with errors in recognition, latency, and inability to be used in a loud or shared space.  
(3) XR headsets are not as commonly available yet, and using a head-mounted display (HMD) might restrict users from freely performing physical movements. 
While our initial prototype does not include an XR interface, given that dance is an embodied activity, we envision future immersive applications to generate choreography in XR. Meanwhile, our current system supports the design of immersive dance experiences. For instance, users can export the 3D dance sequences and import them into an XR scene in Unity~\cite{unity}. This allows a 1:1-sized display that closely resembles the experience of viewing dancers in a dance studio.

\subsection{Functionality} \label{sec:functionality}
\subsubsection{Ideation}
\paragraph{Allows Multimodal Input for Self-expression.} \label{sec:input_modality}
Our system empowers users to input text for choreography generation and harnesses both text and visual data for choreography prototyping. This design choice leverages the synergy achieved by combining textual and visual content to articulate choreographic ideas. Incorporating this design element was driven by insights from our formative study, where participants underscored the importance of integrating verbal communication and visual demonstrations to express their creative ideas. When conveying choreographic ideas through a single mode is inadequate, the two modalities can work in tandem to enhance the expression of their creative vision.

In the case of text input, the text description is transmitted from the user interface to the backend server through a TCP socket~\cite{tcp_server}. Subsequently, it is encoded as text embeddings using CLIP~\cite{radford2021learning} for both dance sequence generation and editing. For video input, the video file is sent to the server and transformed into a 3D dance sequence using VIBE~\cite{kocabas2020vibe}, the same format as the dance sequences generated from text input, for dance sequence prototyping. 

\paragraph{Generates a Variety of Dance Sequences Rapidly.} \label{sec:dance_generation}
To generate dance sequences, users can describe dance movements in the text box, such as ``A person is dancing hip-hop'' and ``A person is doing a moonwalk''. Our system generates three different dance sequences based on each text description and presents the output below the text box as thumbnail images, showcasing the first frame of each sequence using a 2D human skeleton. Users can click the thumbnails to view the corresponding 3D dance sequences in the display area. 
This functionality was motivated by formative study participants who commented that watching dances performed by various artists could expose themselves to diverse styles and help gather inspiration in the process. 
The system allows the generation of dance sequences up to 10 seconds. 
This design choice aligns with findings from prior work~\cite{soga2016body, soga2022experimental}, where short-term dance sequences (around 5 seconds) proved effective for choreographers to improvise, prototype, and edit. Short dance sequences can serve as manageable building blocks to craft new dance pieces and are easy to grasp and manipulate during choreography.

To generate diverse dance sequences based on text, we employed a pre-trained text-to-motion generation model, MDM~\cite{tevet2023human}, as the backbone and fine-tuned it on a large-scale dance dataset AIST++~\cite{Li_2021_ICCV}. The CLIP embeddings of the text input are fed into the fine-tuned MDM for dance sequence generation. Additionally, the converted 3D dance sequences from uploaded videos are used as input for the fine-tuned MDM to allow prototyping. 

\subsubsection{Prototyping}
\paragraph{Enables Extending Dance Sequences.} \label{sec:dance_extension}
The dance extension feature enables users to expand dance sequences by 5 seconds with each user-initiated extension, and this process can be repeated multiple times if desired. The motivation behind this feature stems from the practice of conducting improvisation based on a dance sequence, a prevalent choreography method in dance making~\cite{ciolfi2016choreographers}. By incorporating this feature into our system, we aim to facilitate AI-based improvisation with existing dance data, whether generated by the system or uploaded by users.

To implement this feature, only the 5-second segment to be extended is fed into the fine-tuned MDM for dance generation, while the original segment remains unaltered with masks at each timestamp. Specifically, each timestamp in the original dance sequence is held constant according to a predetermined variance schedule, and the sequence to extend undergoes the standard denoising process of the fine-tuned MDM to generate the extended sequence. The extended segment seamlessly integrates with the original sequence upon generation, resulting in the final longer output. 

\paragraph{Supports Controlling Dance Styles.} \label{sec:style_control}
The style control feature allows users to edit dance sequence styles by applying multiple style transfer options. This feature was motivated by the formative study participants, who highlighted that emotions could inspire new dances and dance could serve as a medium to express emotions. The style options, including angry, childlike, depressed, happy, proud, and strutting, were used by SinMDM~\cite{raab2023single} for motion harmonization. Thus, we used these styles to present a proof-of-concept pipeline to infuse emotional changes into dance movements. 

To implement style control, we generated a reference motion sequence representing each style using the strategy proposed by SinMDM. The style of the reference sequence is then applied to the content of the source dance sequence chosen by users. 
As demonstrated by SinMDM and our experimentation, this approach ensures effective style control while preserving the essential movement details of the source dance sequence. 

\paragraph{Allows Editing Partial Body Movements.} \label{sec:partial_body_edit}
The partial body editing feature facilitates precise and targeted modification of the movements of user-selected body parts. This functionality was motivated by formative study participants with practical experience altering or limiting specific body part movements while permitting the rest of the body to move within defined constraints, thereby inspiring innovative dance movements. This approach prompted us to integrate the partial body editing feature into our system. The edited movements of specific body parts have the potential to influence the overall body movements generated by our system, indicating that novel movements can emerge through the modification of a particular body part. 
Our method for partial body editing shares similarities with the dance sequence extension process. Specifically, we employ a masking technique on the bones of the human skeleton that is not meant to edit and feed the masked human skeleton at each timestamp into the fine-tuned MDM. 

\paragraph{Blends Multiple Dance Sequences.} \label{sec:dance_blend}
The dance blending feature permits users to concatenate any two dance sequences, with the system automatically generating a 5-second sequence for connection. Based on our experiments, choosing a 5-second connecting sequence aims to balance quality considerations, avoiding artifacts in longer sequences and unclear transitions in shorter ones. 
Blending dance sequences is a widely used choreography method to create novel dance sequences by combining distinct styles or genres~\cite{minton2017choreography}. Nonetheless, it presents a non-trivial challenge due to the complexity and diversity of various dance styles. In our system, we utilize generative AI to tackle this challenge. The AI model has been trained on hundreds of data samples spanning multiple dance styles and genres, equipping the model with the capability to blend diverse content effectively and generate novel materials to create choreography. 
We allow dance blending by generating a connecting sequence that combines 2-second prefixes and suffixes from each of the two sequences to be blended. The fine-tuned MDM utilizes the prefixed and suffixed segments as known conditions with masking and generates the intermediate 5-second portion to ensure a smooth transition between the two sequences. 

\subsubsection{Documentation} \label{sec:documentation}
\paragraph{Saves Intermediate Results and the Creation Process.}
The system records the text prompts utilized to generate and edit dance sequences, enabling the tracking of choreographic ideas and creative attempts over time. Additionally, users can add their creations into the \textit{Gallery} to revisit and reuse these intermediate results. 

\paragraph{Exports Final Results in 2D and 3D.}
The system allows users to export the 2D documentation saved in \textit{.mp4} format, presenting animated human skeletons performing the generated dance movements. This format allows easy storage, transfer, and sharing of dance sequences. Users can also download the 3D documentation stored in \textit{.gltf} format, providing them the flexibility to observe dance movements from any perspective with different types of avatar meshes as needed and the ability to further prototype movements in software tools like Blender~\cite{blender}.

\section{User Study}
We conducted a user study to evaluate our system's usability and understand how it may be used in the choreography preparation stage to assist creative exploration and dance sequence refinement.

\subsection{Participants}
To recruit participants for the user study, we posted on the mailing lists of our university's Theater and Dance Department and dance clubs. We specifically sought individuals with extensive choreography experience and selected participants from those who responded to our call and indicated at least three years of experience in the field. Ultimately, we invited six participants (Female: 5, Male: 1), assigning each a participant ID from P01 to P06. The selected participants had choreography experience ranging from four to over ten years. 
To mitigate any bias, we recruited a different set of participants from those involved in the formative study. Since what we learned from the formative study participants informed our system's functionality, we sought users without any preconceived notions about our system to evaluate its usability.
To assess our system as a general-purpose choreography ideation support tool that is designed to handle diverse dance styles, we recruited participants with various choreography backgrounds, allowing us to additionally evaluate the system's genre and style generalizability. Specifically, the participants have contributed to hip-hop, jazz, contemporary, tap, ballet, and gymnastic dance choreography for instruction, commercial dance performances, and online channels. Table~\ref{tab:demographics} summarizes the basic demographic information.

\begin{table}[!ht]
    \centering
    \caption{User study participant demographics.}
    \scalebox{0.9} {
    \begin{tabular}{ccp{0.32\textwidth}}
    \toprule
        ID & Years active & Choreography experience\\
        \midrule
        P01 & 4 & Hip-hop and jazz choreographer and dancer\\
        P02 & 8 & Contemporary dance choreographer\\
        P03 & 5 & Tap dance choreographer\\
        P04 & 10+ & Ballet choreographer\\
        P05 & 10+ & Jazz, contemporary, and gymnastic dance choreographer and dancer\\
        P06 & 5 & Hip-hop choreographer and dancer\\
    \bottomrule
    \end{tabular}
    }
    \Description{This table summarizes the demographic information of user study participants, including their IDs, years of activity, and choreography experiences. Notably, the participants exhibited diverse choreography experiences across multiple dance genres and styles, ranging from at least four years to over ten years.}
    \label{tab:demographics}
\end{table}

\begin{figure*}[!ht]
    \centering
    \begin{subfigure}[b]{0.42\textwidth}
        \includegraphics[width=\textwidth]{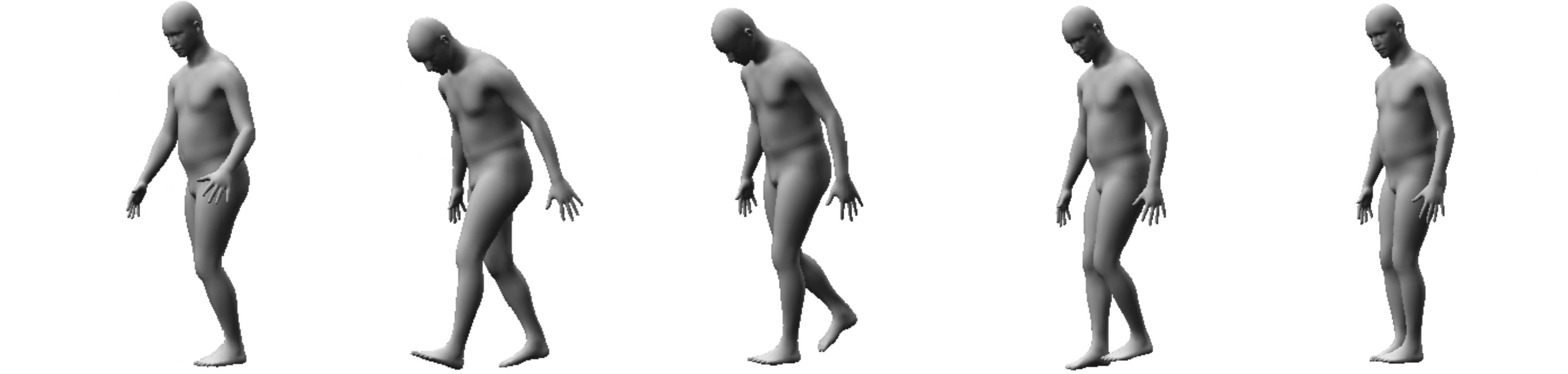}
        \caption{``Raise left heel, keeping only your toes touching the ground. Simultaneously, slide your right foot backward.''}
    \end{subfigure}
    \hspace{0.05\textwidth}
    \begin{subfigure}[b]{0.42\textwidth}
        \includegraphics[width=\textwidth]{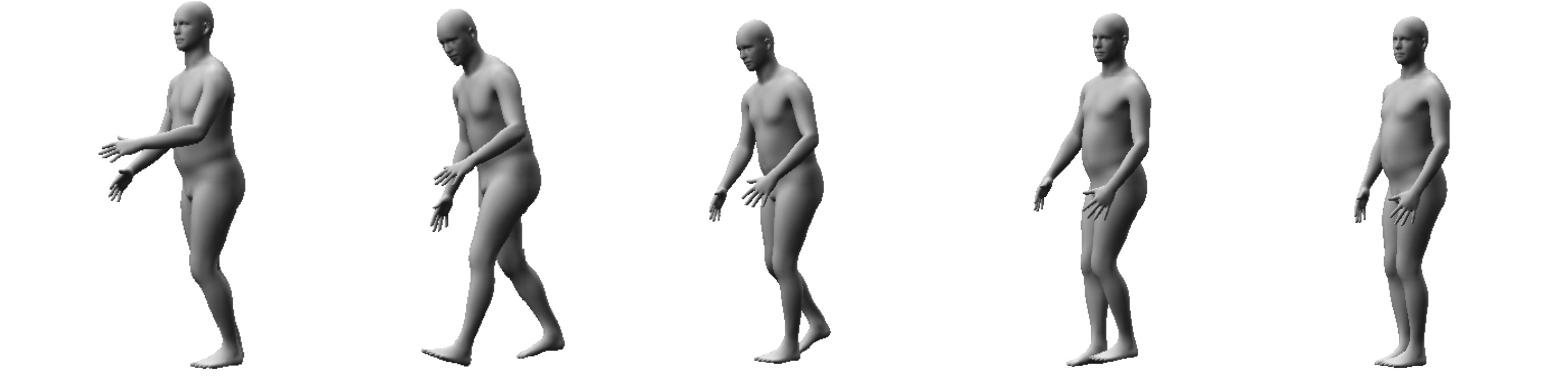}
        \caption{Partial body edit: ``Maintain an upright position with your torso.''}
    \end{subfigure}
    \begin{subfigure}[b]{0.42\textwidth}
        \includegraphics[width=\textwidth]{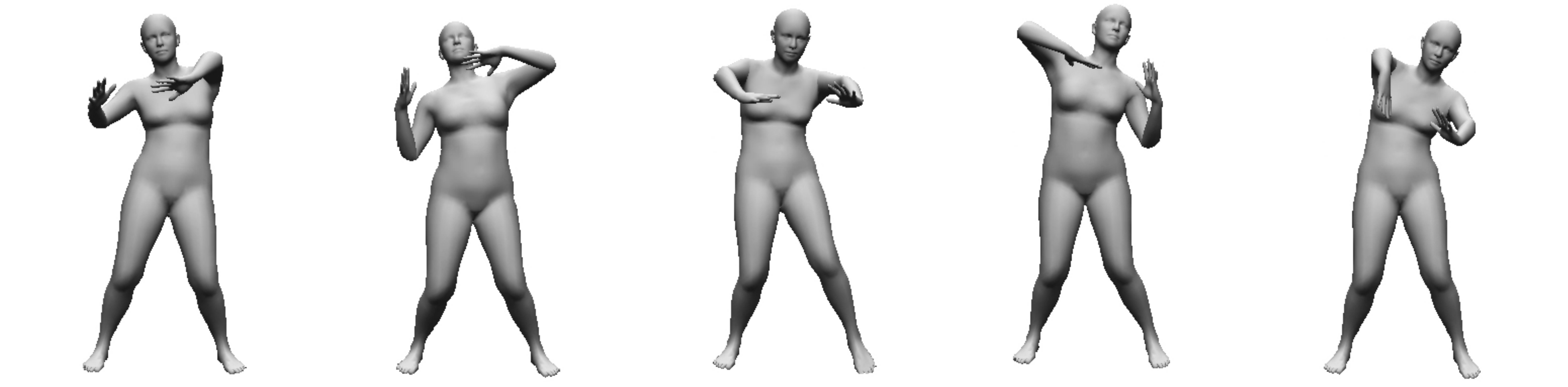}
        \caption{``Chest bumping.''}
    \end{subfigure}
    \hspace{0.05\textwidth}
    \begin{subfigure}[b]{0.42\textwidth}
        \includegraphics[width=\textwidth]{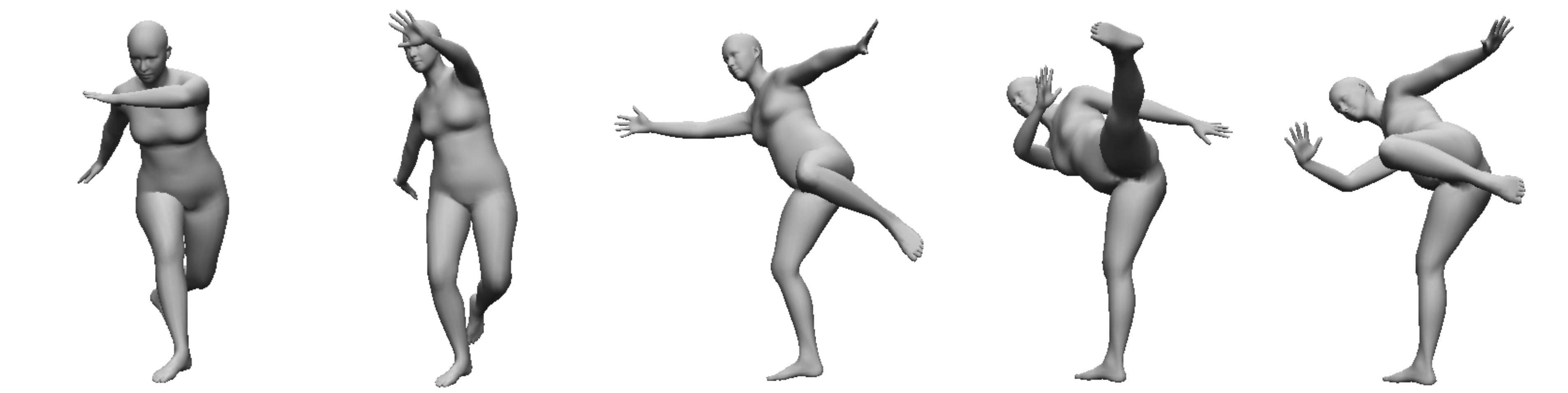}
        \caption{``A person does a hitch kick.''}
    \end{subfigure}
    \begin{subfigure}[b]{0.35\textwidth}
        \includegraphics[width=\textwidth]{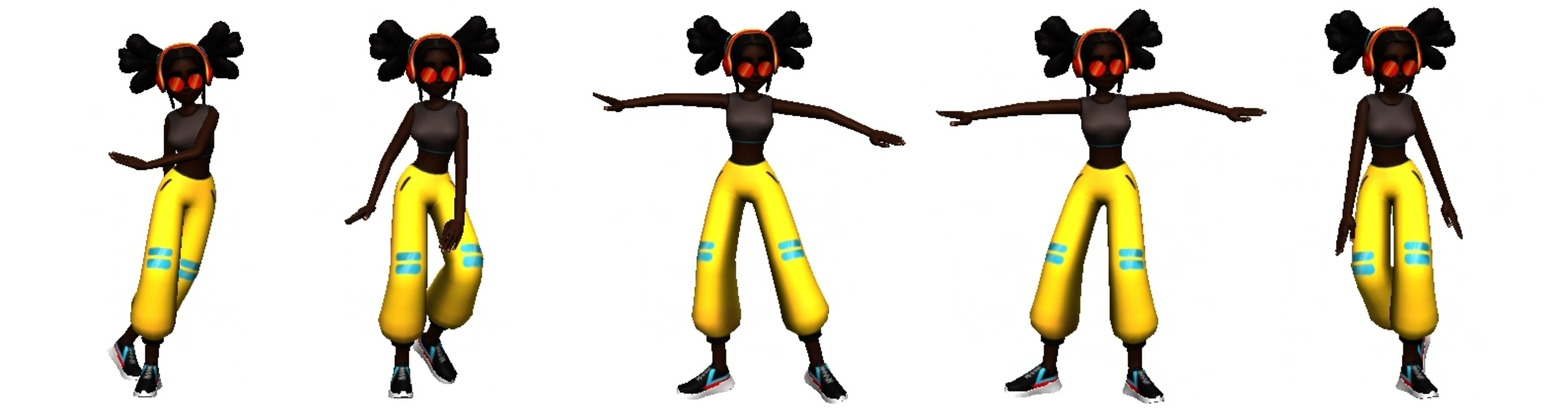}
        \caption{``A person performs a step ball change.''}
    \end{subfigure}
    \begin{subfigure}[b]{0.23\textwidth}
        \includegraphics[width=\textwidth]{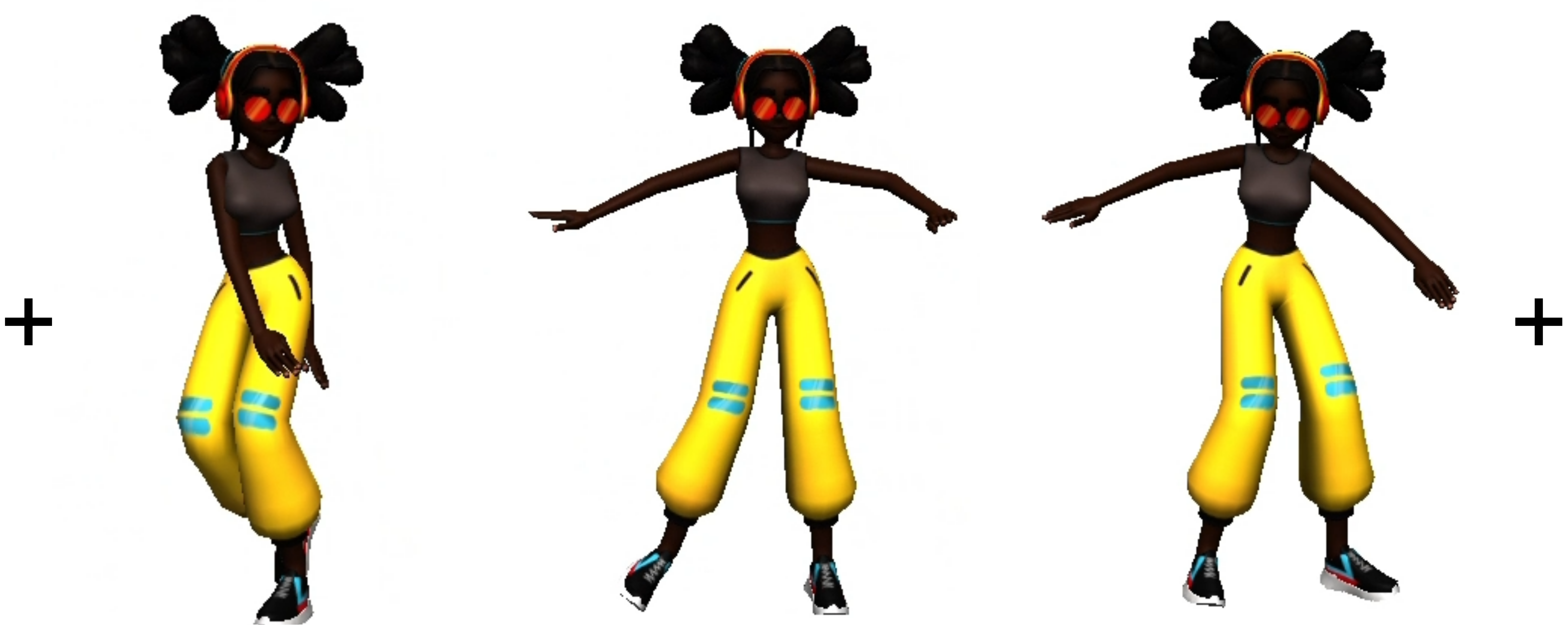}
        \caption{Connecting sequence.}
    \end{subfigure}
    \begin{subfigure}[b]{0.38\textwidth}
        \includegraphics[width=\textwidth]{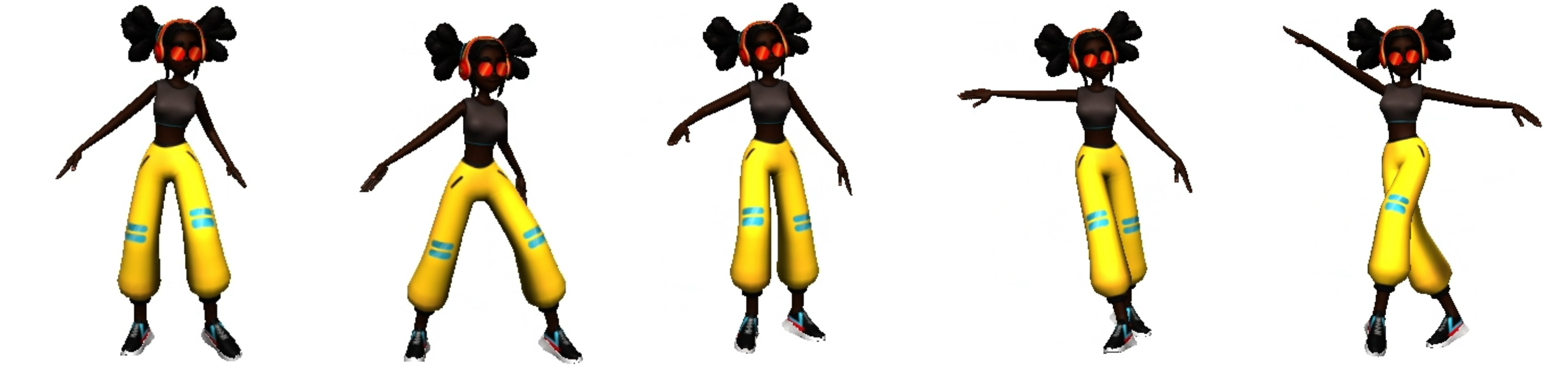}
        \caption{``A person is dancing hip-hop.''}
    \end{subfigure}
    \caption{Results created by user study participants. 
    (a) and (b) Dance sequences are represented by the SMPL male mesh~\cite{SMPL_2015}, generated from ``Raise your left heel, keeping only your toes touching the ground. Simultaneously, slide your right foot backward'', and edited with the partial body condition ``Maintain an upright position with your torso''. 
    (c) and (d) Dance sequences are shown by the SMPL female mesh~\cite{SMPL_2015}, generated from ``Chest bumping'' and ``A person does a hitch kick''. 
    (e), (f), and (g) Dance blending outcomes of ``A person performs a step ball change'' and ``A person is dancing hip-hop'' with a connecting sequence in between, illustrated using the Mixamo mesh~\cite{mixamo}. 
    Each dance sequence is depicted in five frames, while the connecting sequence is presented in three frames.}
    \Description{The figure contains three sets of dance sequences generated by the DanceGen system. The first and second rows of dance sequences are shown by the SMPL male and female meshes, respectively. The third row of the dance sequence is shown by a Mixamo mesh, which is a girl wearing yellow pants and red glasses.}
    \label{fig:generation_results}
\end{figure*}

\subsection{Procedure}
\subsubsection{Consent and Training.}
At the beginning of the study, participants provided us with informed consent. Our local IRB approved the study under protocol \# 21-23-0275. Following informed consent, we conducted an orientation session where we guided the participants through the user interface and explained the system functionality. This instructional phase lasted approximately 5 minutes for each participant. Participants then had 5-8 minutes to familiarize themselves with the system. During this period, they were free to seek clarification, ask questions, and ensure they were comfortable using the system before starting the study.

\subsubsection{Tasks.}
The user study was conducted 1:1 with each participant over Zoom. Participants were given a time limit of 20 minutes to use the system and create dance sequences. There were no constraints on the number, type, or length of dance sequences they needed to create. Participants could initiate the process with any input modality and select any editing options to refine their outcomes. After the study concluded, all participants created at least five dance sequences using the system. They were encouraged to download any results they found appealing. Figure~\ref{fig:generation_results} shows some dance sequences by our participants. 
When the researcher noticed participants omitting a specific feature, we reminded them to explore all system features to ensure comprehensive feedback on all aspects from each participant.

\subsubsection{Semi-structured Interview.}
After completing the tasks, we conducted a semi-structured interview with each participant, which lasted around 20-25 minutes. The interview aimed to gather insights into the participant's (1) experience with the system's functionality; (2) opinions regarding how the system can integrate into their choreography process and the values it provides; (3) feedback on existing limitations and suggestions for future improvements. The complete set of interview questions can be found in Appendix~\ref{sec:user_study_interview}. 

\subsection{Data Collection and Analysis}
During the task session, we took hand-written notes to document participant activities. We made further notes during the interview to record the participants' responses. After completing the study, we performed a reflexive thematic analysis~\cite{braun2006using, braun2019reflecting} on the collected data. This involved reviewing the interview data, coding a subset of the data to extract relevant information, refining the codes, and conceptualizing them into preliminary themes. After the authors had discussions in weekly meetings, a written description of each theme was drafted, followed by another discussion to review and ensure the themes represented the original refined codes. 

\subsubsection{Author Backgrounds.}
To provide context for the perspectives that influenced our interpretation of the formative study, design of the system, and analysis of the user study, we present an overview of the research team's areas of expertise and focus. The authors bring together experience in AI and HCI: Yimeng is a graduate student in computer science and is experienced in developing applied AI systems with a human-centered approach; Misha is an HCI professor in computer science with an interdisciplinary research program that focuses on the design of interfaces, tools, and systems to augment human skills. We have established connections with our university's dance instructors, choreographers, and students to help us understand the practicing artists' needs, which can be supported by our expertise. Our data analysis reveals how AI and HCI researchers viewed the opportunities and challenges of AI-based digital choreography tools based on discussions with practicing choreographers.

\subsection{Results} \label{sec:user_study_results}
Table~\ref{tab:user_study_themes} summarizes the themes obtained from data analysis and a brief description for each theme following the methodology adopted by Lawton et al.~\cite{lawton2023tool}. Based on the data analysis, our results offer insights that extend beyond the system's usability. They reveal the system's potential impact on the broader choreography workflow, informing how it can be effectively used in the preparation stage and thereby, integrated into the choreography process.

\begin{table*}[!ht]
    \centering
    \caption{Five high-level themes developed through our analysis of user study results following the methodology used in~\cite{lawton2023tool}.}
    \scalebox{0.9} {
    \begin{tabular}{cll}
        \toprule
        \S & Theme & Description \\
        \midrule
        \ref{sec:user_study_results_ideation} & \begin{tabular}{@{}p{0.42\textwidth}@{}} \nameref{sec:user_study_results_ideation} \end{tabular} & \begin{tabular}{@{}p{0.51\textwidth}@{}} Participants found that text and video as input allowed them to seek new and fine-tune existing materials. The AI model offered them diverse, novel, and surprising outcomes.
        \end{tabular}\\
        \midrule
        \ref{sec:user_study_results_prototyping} & \begin{tabular}{@{}p{0.42\textwidth}@{}} \nameref{sec:user_study_results_prototyping} \end{tabular} & \begin{tabular}{@{}p{0.51\textwidth}@{}} Participants adjusted their dance sequences in an iterative fashion using the editing options. This process helped them to update results and obtain new ideas.
        \end{tabular}\\
        \midrule
        \ref{sec:user_study_results_documentation} & \begin{tabular}{@{}p{0.42\textwidth}@{}} \nameref{sec:user_study_results_documentation} \end{tabular} & \begin{tabular}{@{}p{0.51\textwidth}@{}} Participants thought that saving intermediate results was beneficial for iteration, and exporting dance sequences was useful for reflection, demonstration, and sharing their creations.
        \end{tabular}\\
        \midrule
        \ref{sec:user_study_results_limitation} & \begin{tabular}{@{}p{0.42\textwidth}@{}} \nameref{sec:user_study_results_limitation} \end{tabular} & \begin{tabular}{@{}p{0.51\textwidth}@{}} Participants reported that some generated dance movements looked like glitches, and gaps sometimes existed between their intents and obtained outcomes.
        \end{tabular}\\
        \midrule
        \ref{sec:user_study_results_reflection} & \begin{tabular}{@{}p{0.42\textwidth}@{}} \nameref{sec:user_study_results_reflection} \end{tabular} & \begin{tabular}{@{}p{0.51\textwidth}@{}} Participants offered both positive and negative feedback on the system functionality and suggested future improvements based on their critiques, including AI output quality and integrated physical prototyping.
        \end{tabular}\\
        \bottomrule
    \end{tabular}
    }
    \Description{This table summarizes the themes obtained from the data analysis of user study results, including section numbers, the identified themes, and concise descriptions of each theme.}
    \label{tab:user_study_themes}
\end{table*}

\subsubsection{Generative AI-driven Exploration in Dance Choreography.} \label{sec:user_study_results_ideation}
Participants found that the system supported the exploration of choreography materials as it, (1) functioned as an on-demand inspiration engine by providing AI-powered resources; (2) expanded their creative horizons through the exploration of diverse dance styles and variations; (3) introduced an element of surprise, potentially leading to unexpected discoveries and innovative choreography.

When using text to describe choreography ideas, participants mentioned that the system was similar to a ``\textit{search engine}'' (P04) that produced outcomes from AI when they felt help was necessary for ideation and brainstorming or when they ``\textit{got stuck}'' (P03). For those lacking preconceived notions, the system offered choreography materials to jumpstart the creative process, as one participant noted: ``\textit{it was helpful to input something into the system and see how AI's responses would be and start from there}'' (P03). As an alternative to online searching, The generative AI system offered participants an approach to acquiring ``\textit{requested resources}'', contrasting to the time-consuming process of browsing through countless dance pieces (P04). This on-demand access potentially saved them significant time previously spent searching for materials (P04). 

Beyond inspiration, the system impressed participants with its ability to produce diverse movement interpretations, mimicking the experience of observing different dancers perform the same piece. This ``\textit{broad scope of materials}'' (P01) allowed choreographers to ``\textit{explore different understandings and styles}'' (P02), fostering a deeper understanding of ``\textit{how AI makes those variations}'' (P02).

\begin{figure*}[!ht]
    \centering
    \begin{subfigure}[b]{0.375\textwidth}
        \includegraphics[width=\textwidth]{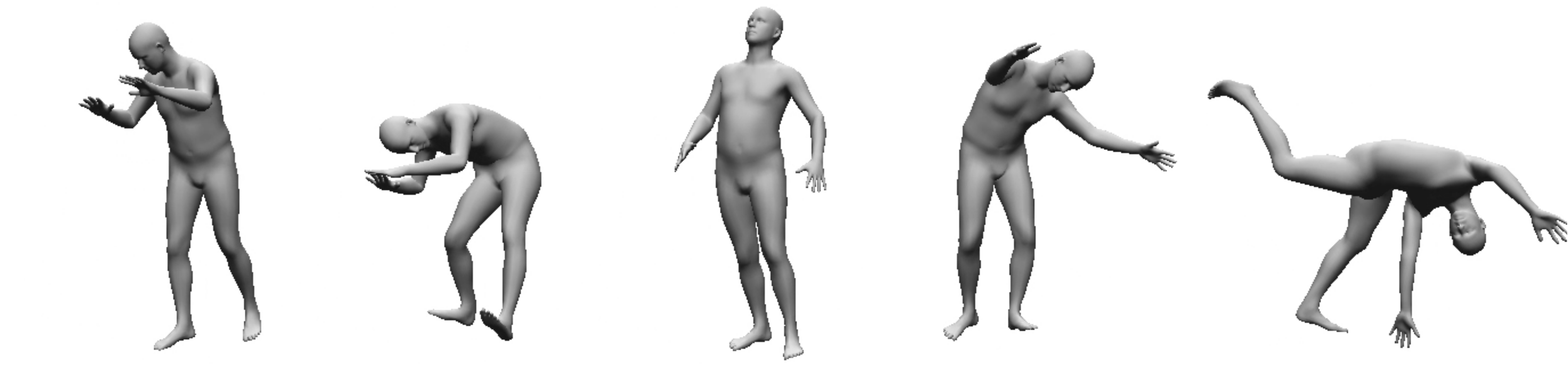}
        \caption{``A person performs a pli\'{e} and then a backflip.''}
        \label{fig:original_sequence}
    \end{subfigure}
    \hspace{0.05\textwidth}
    \begin{subfigure}[b]{0.6\textwidth}
        \includegraphics[width=\textwidth]{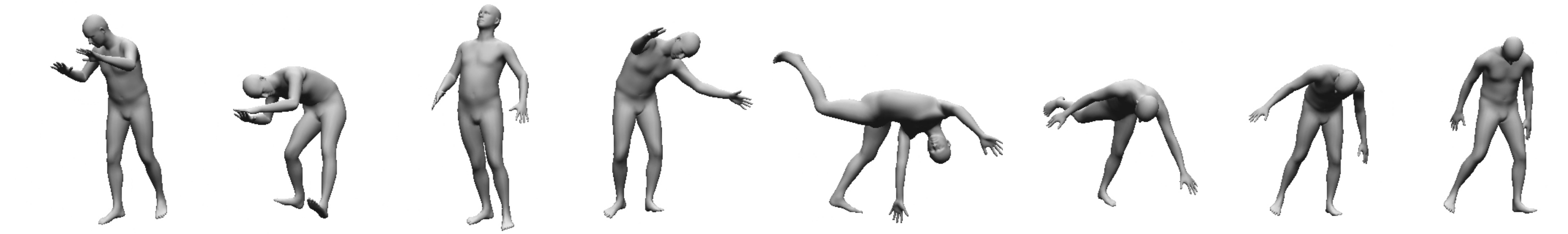}
        \caption{Dance extension.}
        \label{fig:extended_sequence}
    \end{subfigure}
    \hspace{0.05\textwidth}
    \begin{subfigure}[b]{0.375\textwidth}
        \includegraphics[width=\textwidth]{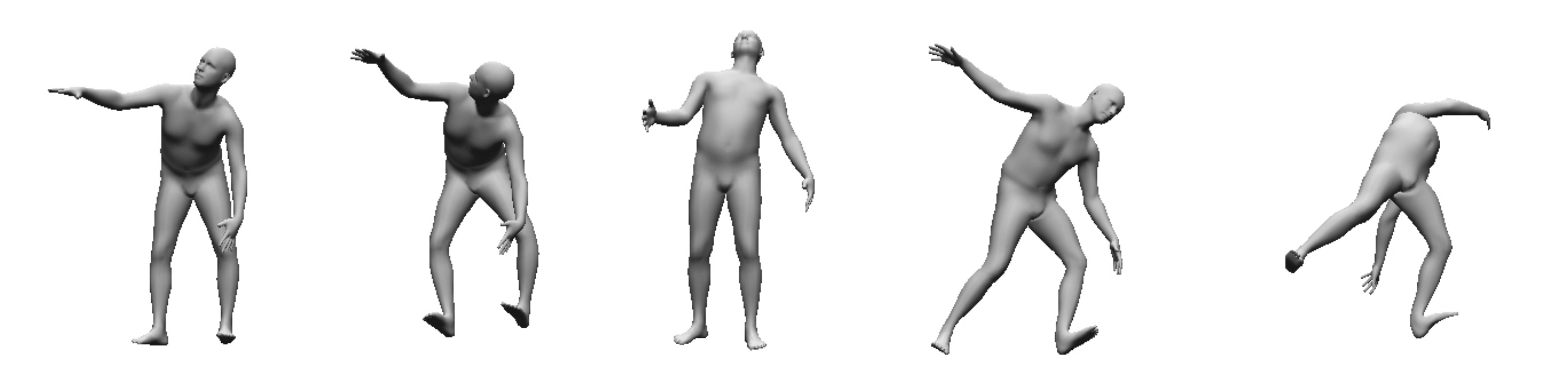}
        \caption{Style control: strutting.}
        \label{fig:style_control_angry}
    \end{subfigure}
    \hspace{0.05\textwidth}
    \begin{subfigure}[b]{0.375\textwidth}
        \includegraphics[width=\textwidth]{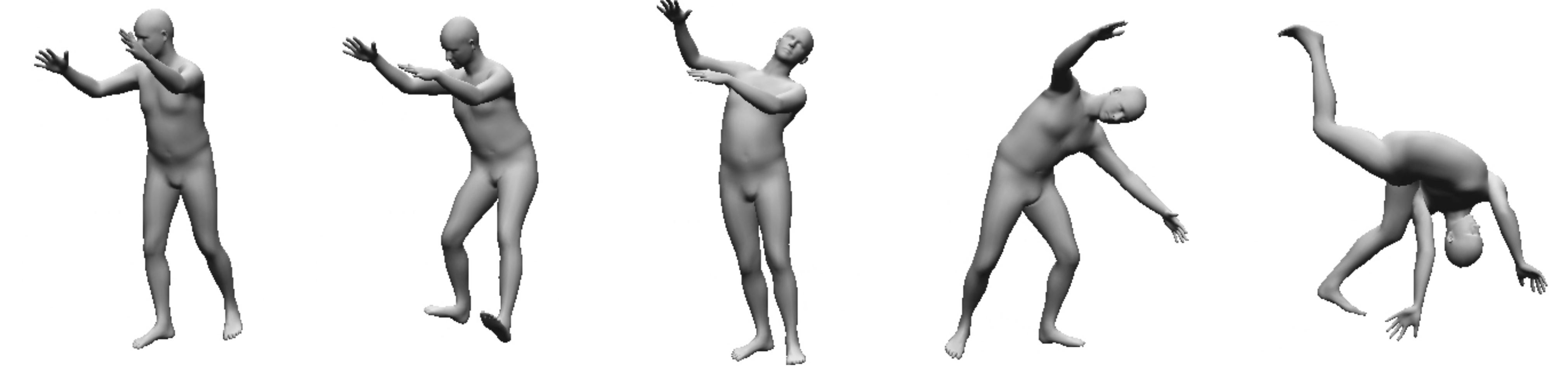}
        \caption{Partial body edit: ``Keep the arms raised.''}
        \label{fig:upperbody_armsup}
    \end{subfigure}
    \begin{subfigure}[b]{0.37\textwidth}
        \includegraphics[width=\textwidth]{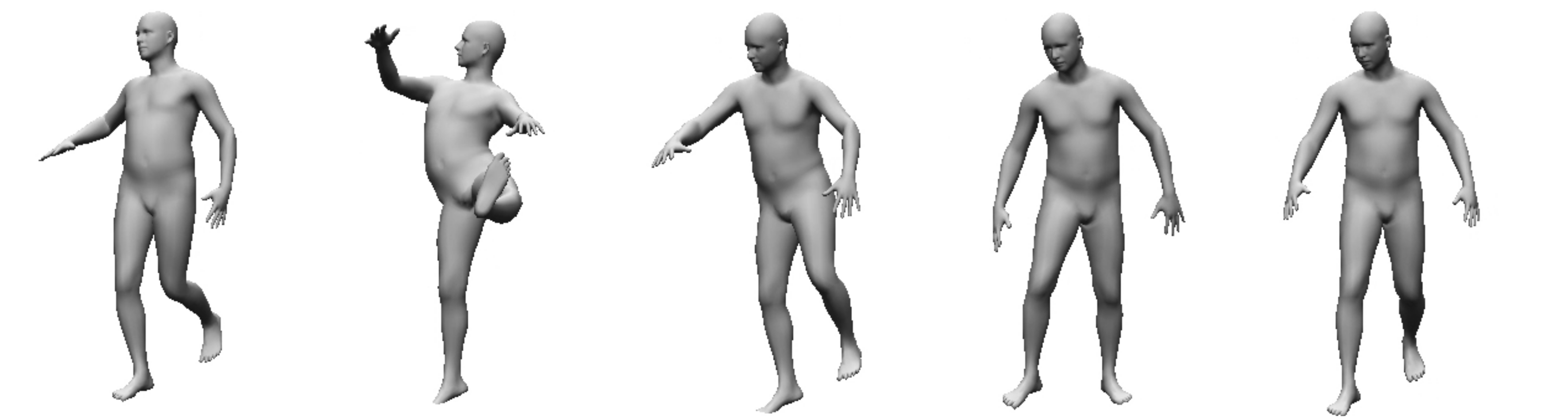}
        \caption{``A person performs a kick ball change.''}
        \label{fig:blend_original_sequence}
    \end{subfigure}
    \begin{subfigure}[b]{0.23\textwidth}
        \includegraphics[width=\textwidth]{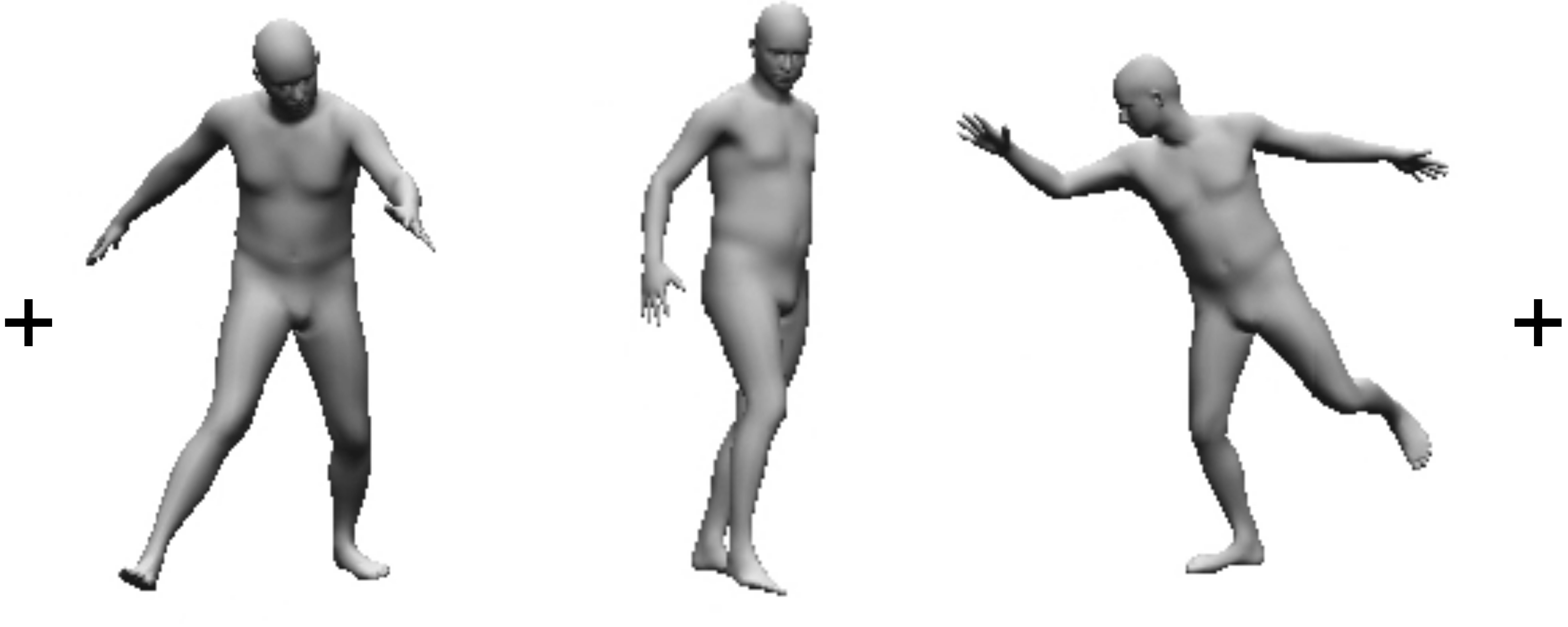}
        \caption{Connecting sequence.}
        \label{fig:blend_connect_sequence}
    \end{subfigure}
    \begin{subfigure}[b]{0.375\textwidth}
        \includegraphics[width=\textwidth]{plie_backflip_combo.png}
        \caption{``A person performs a pli\'{e} and then a backflip.''}
        \label{fig:blend_second_sequence}
    \end{subfigure}
    \caption{Dance generation and editing results. 
    (a) Generated dance sequence shown as five frames based on ``A person performs a pli\'{e} and then a backflip''. 
    (b) The original dance sequence is extended by 5 seconds, with the extended portion depicted by three frames. 
    (c) Style control, strutting, is applied to the original dance sequence. 
    (d) Partial body editing based on ``Keep the arms raised''. 
    (e), (f), and (g) A dance sequence generated from ``A person performs a kick ball change'' is seamlessly blended with the original dance sequence through a 5-second connecting sequence shown as three frames.}
    \Description{The results are shown using the SMPL male mesh performing the system-generated dance movements.}
    \label{fig:editing_results}
\end{figure*}

More importantly, the system introduced an element of surprise and serendipity into the creative process. It could produce content that participants would not have conceived of on their own, pushing them to ``\textit{break comfort zones during ideation}'' (P06). Figure~\ref{fig:editing_results} showcases some of these unexpected combinations, such as the juxtaposition of a classical ballet pli\'{e} followed by a contemporary backflip. These surprising outputs could serve as stimuli for innovative and unconventional choreography. 

\subsubsection{Iterative Editing for Dance Refinement, Understanding, and Discovery.} \label{sec:user_study_results_prototyping}
Participants remarked that iterative editing movements allowed them to refine their work in powerful ways as it, (1) provided an approach for tailoring dance sequences to achieve desired outcomes; (2) facilitated a deeper understanding of the relationship between user input and AI output; (3) fostered the exploration of new creative directions and the discovery of fresh movement possibilities. Figure~\ref{fig:editing_results} presents a set of edited results using the editing options of our system. 

The system's editing features appeared to be the cornerstone of participants' creative process by allowing for adjustments to support a workflow of iterative refinement. For instance, participants commented that ``\textit{the editing options allowed me to change what I didn't like or I thought could be improved}'' (P01). The system's fast generation capabilities facilitated rapid prototyping, allowing participants to ``\textit{get prepared with materials and early tests for choreography}'' (P04). The iterative editing approach proved particularly beneficial for challenging or unfamiliar choreography. For example, participants noted that iteration would not take too much effort for familiar styles but becomes crucial for tackling new territory (P05).

For participants with experience in generative AI tools like DALL-E, DanceGen offered a distinct advantage. They appreciated the system allowed them to see how their controls affected AI outputs, with comments like ``\textit{I feel I have control over what I want to make by updating AI's results ... I kinda have a better clue of how my inputs influence the outcome}'' (P02). Unlike purely automatic generation tools, DanceGen's functionality fostered an iterative approach, which could be likened to collaboratively working on a piece with another choreographer, modifying and evolving creative ideas through discussion.

The ability to prototype dance sequences through editing opened doors to new ideas and materials. Partial body editing and style control resonated strongly with participants, aligning with established dance-making practices identified in the formative study. Comments like ``\textit{I could see how the rest of the body movements changed to coordinate the partial body I edited}'' (P01) illustrated how the partial body editing feature fostered the generation of new movements according to specified body movement constraints. Similarly, the system's ability to adjust styles, as noted by P04 who observed that ``\textit{the original dance moves changed apparently to reflect the happy and angry styles}'', showcased how AI can mimic aspects of a choreographer's strategies to incorporate a variety of emotions for creating new movement sequences.

\subsubsection{Documentation as a Bridge for Iterative Creation and Collaborative Workflow.} \label{sec:user_study_results_documentation}
Participants commented that the system's documentation technique supported seamless integration into the creative process as it, (1) enabled a loop for revisiting, refining, and building upon past work within the system; (2) bridged the gap between individual creation and collaborative workflows through shareable documentation.

Participants commented that the \textit{Gallery} served as a central hub, allowing them to ``\textit{retrospect, redo, and reuse}'' (P01) their creations. This ``\textit{retrospect}'' feature, as described by P01, enabled participants to revisit past iterations and build upon them, fostering a continuous improvement cycle. Furthermore, the 3D animated avatar proved helpful for choreography prototyping. Participants endorsed its ability to provide ``\textit{multi-perspective}'' views (P06), akin to observing a dancer in a real working space (P06). 

Moreover, the system's downloading feature addressed the need for long-term access and collaboration. Participants could export their creations for future use, as noted by P05: ``\textit{I can save results ... to revisit and for others to view to collaborate with them.}'' This eliminated the need for repeated demonstrations and ensured consistency when sharing with collaborators. An additional benefit emerged, as remarked by P02, that the downloaded 3D documentation proved compatible to use in other software like Blender. This compatibility can open doors for further iterations using participants' preferred software tools (P02), extending the creative potential beyond the initial creation within DanceGen.

\subsubsection{Limitations in AI Fidelity and Intent Understanding Hindered User Experience.} \label{sec:user_study_results_limitation}
Participants identified limitations in the AI model that impacted their experience, including (1) occasionally producing unrealistic movements and styles that did not align with their expectations; (2) struggling to grasp specific user intentions, leading to mismatched outputs.

Although participants found unusual dance movements generated by the system helpful in stimulating new thinking, some participants treated these results as glitches produced by the AI model. For example, P06 perceived the AI-generated movements as ``\textit{unnatural and rigid}''. Additionally, the system struggled with specific dance genres, with comments like ``\textit{ballet was not very similar to human dancing}'' (P04). Such artifacts in generated pli\'{e} as shown in Figure~\ref{fig:editing_results} present its limitations in replicating certain styles. 

Additionally, gaps sometimes existed between participants' intentions and the system's output. Participants reported instances where the system misinterpreted dance verbs and terms used in text prompts, leading to undesired results, as noted by P01: ``\textit{the system interpreted `shuffle' as something else.}'' While iterative refinement could sometimes address these misinterpretations, the system occasionally produced outputs that deviated further from expectations. For example, P04 described applying the ``\textit{depressed}'' style and finding the result to be a confusing mix of ``\textit{depressed}'' and ``\textit{angry}'' even after multiple tries. These inconsistencies could cause user frustration and limit the system's effectiveness in achieving the desired creative vision.

\subsubsection{Balancing Potential and Limitations --- Towards AI-powered Exploration and Physical Realization.} \label{sec:user_study_results_reflection}
The user evaluation indicates a balance of positive and negative experiences with DanceGen. The system offered promising potential as a tool for creative exploration, expansion of movement possibilities, and iterative digital prototyping. On the other hand, further development in AI model capabilities is needed to better translate user intent into desired dance movement, and improvement of the system design to integrate physical prototyping is desired to bridge the gap between digital and physical dance creation.

Participants' positive feedback centered on the system's ability to spark creative ideation. As P06 noted, ``\textit{AI can generate dance movements that I wasn't able to do or haven't ever thought of. When I want to make something new, I am happy to try dance movements that are unfamiliar to me.}'' This ability to introduce uncommon movements resonated with participants seeking fresh inspiration. Furthermore, participants found that DanceGen could seamlessly integrate into their existing creative process. P04's comment, ``\textit{it aligns with what I usually do to make dance iteratively}'', highlighted how the system supported the iterative nature of choreography creation. The potential usages of DanceGen envisioned by participants included modifying existing pieces, overcoming creative roadblocks, creating dance specifically for digital use, and blending human-made and AI-generated choreography.

However, valuable critiques also emerged. Some participants expressed that the system output did not always make sense to them or match their needs, such as ``\textit{some movements looked similar across different [generated] samples. I want to know how the variations were built}'' (P04), and ``\textit{I would prefer the generated outcome to match what I described and edited as closely as possible, but this system couldn't guarantee it}'' (P06). Participants also commented that simple interaction could help to make editing and refining more effortless. For instance, ``\textit{allowing dragging the virtual character to edit movements, like dragging the joints would be an easier way to interact with a digital tool}'' (P02). 

Looking beyond the digital realm, participants emphasized the importance of integrating physical prototyping into the creative process. Comments like ``\textit{physical prototyping needs to follow digital prototyping using the system}'' (P04) underscored DanceGen's potential as a tool for preparing materials for dance creation and refinement in a studio. Furthermore, participants recognized the limitations of a digital avatar in conveying the nuanced elements of dance, such as ``\textit{emotion and energy}'' (P03). While the user interface offered emotion modifiers, participant feedback suggested a need for further development in this area. Ideally, future AI models should be able to better integrate these implicit aspects into the generated content.

\section{Discussion}
The user study offered us insights into the benefits and limitations of using the system to facilitate the choreography preparation stage. In this section, we discuss our analysis of the user study results and lessons learned from the design and evaluation of DanceGen. 

\subsection{Describe and Visualize Choreography Creation Ideas}
Effectively translating creative concepts from the mind into physical movements that convey the choreographer's creative vision is a fundamental aspect of choreography creation~\cite{calvert1989composition}. Physical demonstration by choreographers themselves is a prevalent method to express and visualize choreographic ideas. However, potential fatigue may arise from continually watching oneself in mirrors during dance creation, a comparable effect to Zoom fatigue, where constant real-time self-observation during video chats is considered tiring~\cite{bailenson2021nonverbal}. To alleviate this, choreographers often invite other dancers to test and prototype choreographic ideas. This approach can also allow choreographers to view their choreographic ideas from a third-person perspective, see the motions in a spatial context, and receive feedback as others may translate ideas into dance movements based on their unique understanding and experience. 
In our user evaluation, participants found that the AI-generated outcomes gave them an AI perspective on translating choreographic ideas into dance movements. The animated digital avatars offered an alternative format for third-person demonstration without the fatigue associated with continuous self-observation in mirrors. Additionally, the digital avatars enabled participants to observe the evolution of dance sequences throughout the choreography process. As the digital record accumulates over time, it can serve as a database to reflect each choreographer's style --- akin to the concept of the choreographer's knowledge base identified in~\cite{calvert1989composition} and facilitate material reuse for their continuous creative endeavors. 

Converting choreographic ideas into physical movements emphasizes the kinematic aspect of choreography creation. However, the non-kinematic aspect is also indispensable in expressing and visualizing the choreographer's creative vision~\cite{zhou2021dance}. Laban Movement Analysis (LMA)~\cite{laban1975modern} has identified two components within the non-kinematic aspect: \textit{effort} --- the movement quality, and \textit{shape} --- the way and motivation the body changes its shape. In our user evaluation, participants expressed the need for an improved approach to convey ``emotion and energy'' (Section~\ref{sec:user_study_results_reflection}), which exemplified the LMA effort component. Capturing the non-kinematic aspect is non-trivial due to its imprecise and abstract nature, making it difficult to incorporate into a choreography-support system compared with quantifiable features like the human skeleton movement. Previous attempts have utilized Kinect sensors~\cite{ran2015multitask} and Inertial Movement Units and electromyography sensors~\cite{camurri2016dancer} to model and analyze dance movement quality. However, using computational methods, including AI-based approaches, to analyze and model the non-kinematic aspect of human movement and generate new movements based on adjustable LMA effort and shape parameters for choreography support remains underexplored. This poses an exciting research area since computational approaches can potentially enhance the efficiency of choreography creation by minimizing dependence on human embodied perception~\cite{alaoui2017seeing} and augment the affectivity and expressiveness of generated movements by taking the non-kinematic aspect into account~\cite{samadani2020affective}. 

\subsection{Spark New Ideas for Choreography}
Our participants recognized some outcomes generated by DanceGen as unusual and expressed both positive and negative opinions regarding these AI-generated artifacts (Sections~\ref{sec:user_study_results_ideation} and \ref{sec:user_study_results_limitation}). These artifacts can stem from machine hallucination and the absence of strong physical constraints on joint motions. 
The positive feedback aligned with prior work that utilized simulated unusual movements to stimulate novel ideas during dance improvisation~\cite{soga2016body, soga2022experimental}. 
To mitigate the negative effects pointed out by some participants, a possible solution is to implement a robust physical guide~\cite{yuan2023physdiff, maeda2022motionaug} or re-rig the avatar mesh with a biomechanics skeleton, such as SKEL~\cite{keller2023skel}, to prevent AI-generated artifacts like odd joint twisting, gain/loss of volume, and foot sliding. While this approach can help prevent unnaturalness and enhance realism, the results may tend to resemble widely available human-made dance movements, potentially diminishing the unique contribution offered by AI. 

\begin{figure}[!ht]
    \centering
    \begin{subfigure}[t]{0.23\textwidth}
        \includegraphics[width=\textwidth]{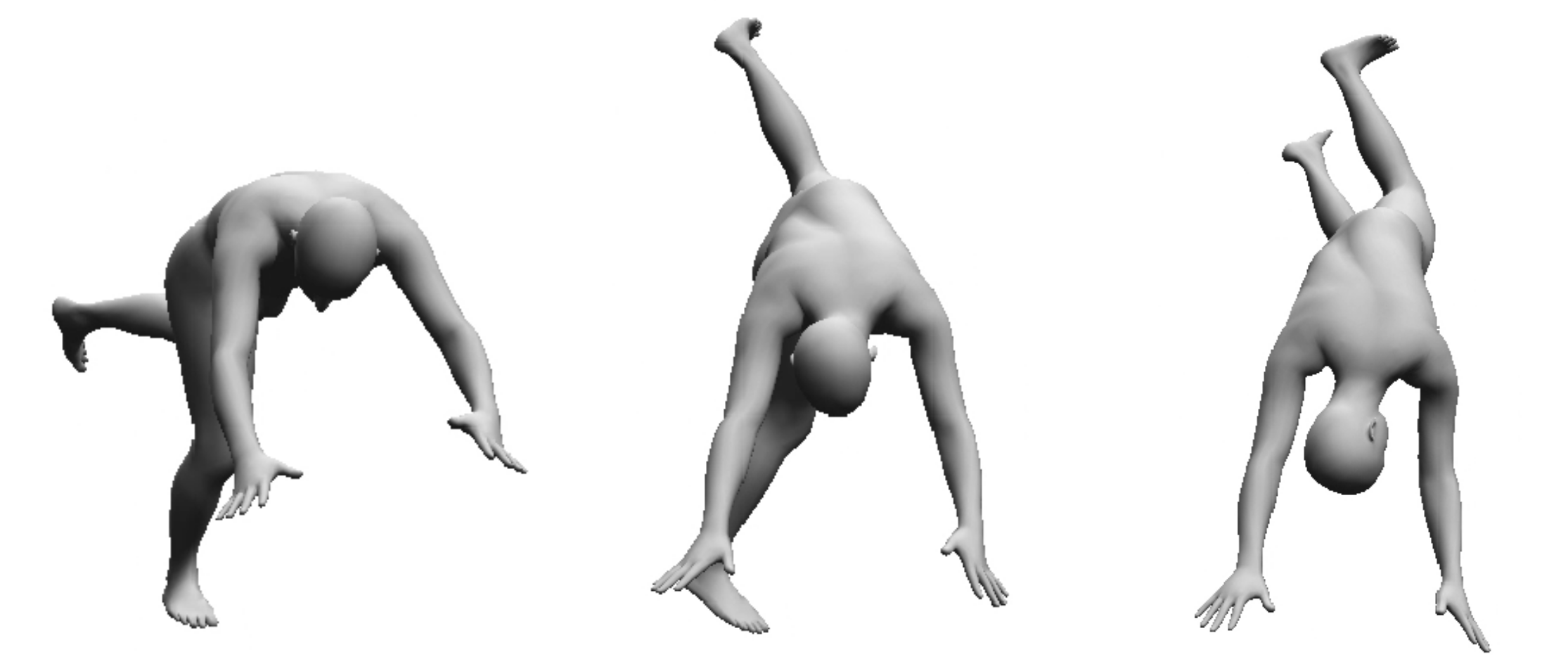}
        \caption{``Handstand.''}
    \end{subfigure}
    \hspace{0.05\textwidth}
    \begin{subfigure}[t]{0.25\textwidth}
        \includegraphics[width=\textwidth]{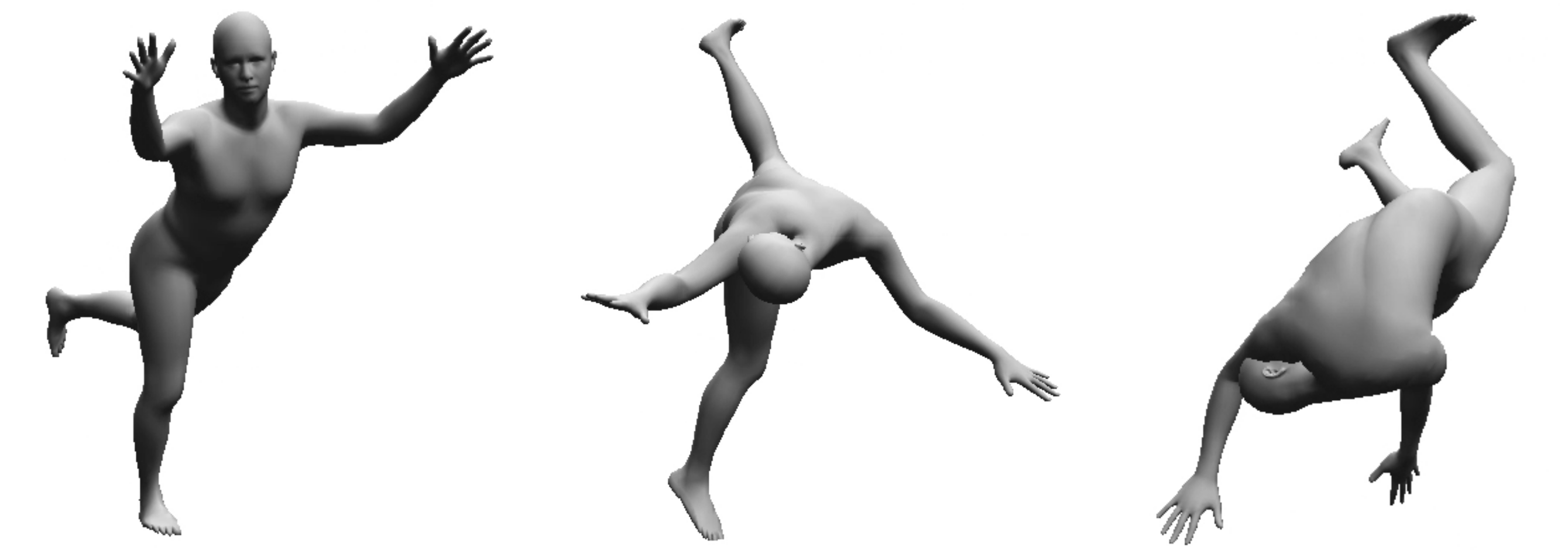}
        \caption{Partial body edit: ``Keep both arms raised up.''}
    \end{subfigure}
    \caption{(a) Dance sequence generated by ``Handstand''. (b) Edited dance sequence based on (a) with the partial body condition ```Keep both arms raised up''. Each dance sequence is displayed in three frames.}
    \Description{The results are shown using the SMPL male mesh performing the system-generated dance movements.}
    \label{fig:handstand_results}
\end{figure}

Similar artifacts can arise from sources other than generative AI models. For instance, VR device tracking errors can lead to unusual arm and leg movements in the tracked human skeleton. While tracking errors may resemble some AI-generated artifacts, whether choreographers perceive these errors as helpful for ideation remains an underexplored aspect. Artifacts generated by AI or errors caused by tracking methods present a promising research question regarding how creativity may emerge from technology glitches. For example, our system can generate movements like ``handstand'' with the partial body condition ``keep both arms raised up'', see Figure~\ref{fig:handstand_results}. In this case, the generated motion sequence features both arms and legs raised when doing a handstand, which is challenging for a human to perform without a prop. However, looking at it from a different perspective, two people could collaboratively achieve this movement, one raising their arms and the other their legs, an idea that could contribute to creating a duet dance from solo ideation --- new ideas may strike when choreographers view errors or impossibilities from different angles. 
Embracing unusualness shares a similar motivation with adopting improvisation to enhance creativity, as both aim to encourage choreographers to distance themselves from habits and develop new perspectives~\cite{gerber2007improvisation}. Our evaluation demonstrated that AI had the potential to offer a fresh angle, distinct from human perspectives, and stimulate new thinking in choreography creation.

Zooming into the entire choreography process, creativity is crucial in multiple iteratively occurring stages (Figure~\ref{fig:choreography_process}). 
This demands both convergent thinking, which involves evaluating dance sequences and choreographic ideas and making decisions, and divergent thinking, which entails devising novel ideas and staying creative over an extended period~\cite{tversky2011creativity}. Divergent thinking is a difficult task because it is ``information hungry''~\cite {wang2023popblends, veale2019conceptual}. AI-supported systems can help with divergent thinking in various ways, including their tireless responsiveness for iterative ideation and prototyping and their ability to generate diverse and unseen content based on high-level concepts extracted from extensive datasets on which they were trained. 
Incorporating the strengths of AI may lead to the design of more effective choreography-support systems that are well-aligned with the existing choreography process. This integration may offer opportunities for the next generation of choreography tools to enhance creativity. While the willingness to incorporate AI into workflows depends on personal experience and preferences, our user study participants expressed enthusiasm for newer tools to support their workflows.

\subsection{Integrate Physical Prototyping into Choreography Creation}
The concept of using technology to support choreography creation dates back to the 1990s with systems like Life Forms~\cite{calvert1989composition}. This early work has emphasized the importance of addressing time-consuming physical needs in choreography and introduced a computer-based choreographic software tool to assist in this regard. In the past few decades, researchers have worked on building tools to support physical activities during choreography by extending the human body expressivity with computing (Section~\ref{sec:related_work}). Today, AI can become a powerful computational backend to aid physical needs through digital prototyping, offering clear advantages as evidenced by our user evaluation (Section~\ref{sec:user_study_results_prototyping}). 
Using AI to support digital prototyping aligns with the longstanding recognition of the significance of addressing physical demands in the choreography process and also empowers those who may be physically limited due to injury or disease by providing an accessible means of continuing to generate and explore dance motion ideas digitally. 

While the DanceGen system highlighted the advantages of digital prototyping, our study participants reflected on the necessity of physical prototyping in the choreography workflow (Section~\ref{sec:user_study_results_reflection}).
This finding echoed earlier research~\cite{calvert1989composition} underscoring that physical demonstrations by live dancers are required to capture and represent ``energy flow'', ``emotion flow'', ``essence'', or ``aesthetic shape''. 
To meet the need for physical prototyping of dance movements and reduce dependence on time-consuming physical prototyping, an efficient embodied prototyping approach is demanded. 
Our participants suggested that physical prototyping with the documented choreographic materials could complement our digital system. This approach may assist choreographers in physically adjusting movements and transferring those modifications back to the digital system to update dance sequences. Integrating physical activities and digital prototyping can become particularly valuable when digital prototyping alone cannot fulfill all prototyping needs, such as conveying implicit information through motion. Achieving this integration requires a technique that spans multiple stages of the choreography process. This includes the preparation stage, emphasizing ideation, and the studio stage, focusing on physical prototyping. While our system allows participants to upload dance videos as a physical demonstration of movements, participant feedback on physicality indicates the desire for a real-time motion capture mechanism to allow them to physically ``edit'' dance sequences. To this end, spatial display and interaction technologies could be leveraged to provide a seamless transition between the digital system and physical actions by allowing either outside-in or inside-out body tracking and motion capture or recording a video in real time with 3D pose estimation. As discussed in Section~\ref{sec:ui_platform}, at present, DanceGen can be extended to spatial interfaces, such as XR, by transferring the generated 3D content to an XR frontend. Such spatial display and interaction techniques may further inspire novel ideas and provide unique and enjoyable experiences for choreographers~\cite{long2017designing, jochum2019tonight}. These advantages contribute to the overall benefits that spatial techniques may bring to the choreography process, improving not only prototyping approaches but also ideation strategies and providing an engaging user experience. 
It is also essential to consider the design of flexible and simple interaction, such as adopting verbal communication and example-based manipulation. These aspects, preferred and suggested by our participants, may contribute to an unobtrusive experience and support the seamless combination of digital and physical elements in the choreography process.

\section{Limitations and Future Work}
\subsection{System Design}
Prior research~\cite{rezwana2022designing} has classified AI-assisted computational systems into those creating novel artifacts, supporting human creativity, or allowing human-AI co-creativity. According to our user evaluation, our system falls into the first two categories, as it can facilitate the generation of new dance sequences and enhance choreographers' creativity by assisting them in shaping creative ideas through generated outcomes. The current role of AI in the choreography process resembles that of a tool that reacts to human needs, as opposed to a collaborator that offers suggestions and contributes to the shared product. Previous work has shown that the collaborative production of creative outcomes tends to be more innovative than what individuals can achieve alone~\cite{sawyer2009distributed} and suggested collaboration and interaction paradigms for human-AI choreography co-creation~\cite{liu2024interaction}. We plan to investigate how this work can be extended to allow AI models to offer feedback based on user activities and contribute to intermediate and final outcomes, fostering creativity through collaborative effort.

\subsection{Technical Implementation}
Our system supports general-purpose choreography creation using a generative AI-based system, which cannot guarantee an alignment with the distinct requirements of choreographers. Choreographers who work with different dance genres often use heterogeneous descriptions to articulate their ideas and needs. Meeting these diverse requirements using a single AI system is challenging, mainly due to the limited scope of dance movement descriptions in the AI model's training data. Any descriptions beyond this scope may result in random outcomes hallucinated by the AI model. This poses an important research problem for tools designed to support experts vs. amateurs --- professional vocabulary and terms that might be unfamiliar to the general public are often common and basic within a specific domain. To address this problem partially, we introduced the option for users to input videos, allowing them to demonstrate dance movements unknown to the AI model visually. To tackle this issue more comprehensively, future work is needed to develop customized and tailored dance generation algorithms by building new datasets or training on dance datasets with appropriate textual and visual labels specific to different dance genres and styles. 

\subsection{User Study}
Although our user study offered valuable insights into the design and impact of AI choreography-support systems, the small sample size, potential gender bias among participants, and limited range of choreography genres represented may restrict the generalizability of the findings. We plan to involve a larger and more diverse group of choreographers to address this limitation for a more comprehensive evaluation. 
Additionally, the study prioritized system usability, evaluating its effectiveness in supporting creative exploration and digital prototyping for choreography preparation. Participants successfully generated and refined multiple dance sequences within the 20-minute session, demonstrating the system's potential to streamline early-stage choreography development. This study setup can be seen as a condensed, digital version of the preparation process but specifically focused on dance movement ideation and early prototyping. In a typical preparation stage, ideation can take hours or even days to create a complete piece for dance instruction or performance, according to our formative study participants. They need to gather inspiration from external resources and their own experiences and identify promising ideas. Based on the chosen set of materials, early testing and refinement can take many more hours as choreographers filter out unsatisfactory concepts and iteratively polish the piece. The extensive time investment in choreography preparation can be stressful, especially given time and budget constraints. To shorten the time spent searching for inspiration, artists can use apps with advanced recommendation algorithms, such as TikTok, that suggest related materials more efficiently based on browsing history and preferences compared with online keyword-based searching. However, subsequent choreographic prototyping using the found materials is not supported by such apps and remains desirable based on formative study participant feedback. 
Our system is designed to address some of the challenges that exist in the preparation stage by enabling efficient new movement generation, faster digital prototyping, and documentation for iterative creation and future use. While our AI-based system aims to enhance efficiency, it is uncertain whether it will effectively reduce the time spent during the preparation stage. This is because a choreographer might invest an equivalent amount of time in generating and modifying dance movements using our system as they would when browsing and watching YouTube videos for inspiration.

Participants in our user study considered our system's applicability at various stages during ideation. This included scenarios where they lacked specific requirements and aimed to discover the AI's capabilities, as well as situations where they had an initial concept but sought further creative exploration. Given the varied potential usage scenarios, the manner in which individual choreographers use the system may result in diverse outcomes regarding its usability, overall efficacy, and integration within their current choreography creation process.

The limitations of the in-lab user study restricted a comprehensive analysis of the system's integration into the preparation stage, with potential downstream impact on other stages, including studio rehearsals, dance performances, and post-performance reflection. 
While the usability evaluation has provided valuable insights into the system's functional efficacy and how participants thought the system could support their creative endeavors, further assessments such as extended workshops or multi-week usage with specific design goals such as creating short videos or live performances by one or more dancers, could facilitate a more comprehensive understanding of how the system might support different types of existing choreography design practices. Such evaluations would enable us to gather more in-depth user experiences, encompassing the entire choreography life cycle and offer richer insights for future development.

\section{Conclusion}
This paper introduced DanceGen, a generative AI-powered system designed to assist choreographers in ideation and prototyping in the choreography preparation stage. To build this system, we invited participants with choreography experience for a formative study to help us develop design specifications for our system. The system features a web-based interface and a diffusion model-supported backend, enabling fast generation, iterative digital prototyping, and documentation of dance sequences. An expert user evaluation was performed to assess the system's usability. The evaluation results offered us insights into the system's functionality and potential use to facilitate the choreography preparation stage. Based on these findings, we identified key areas for future design and development in AI choreography-support systems, including facilitating creative idea description, sparking inspiration by unusual movement, and integrating with physical prototyping. These insights can be used to guide the continued development of DanceGen and similar systems.

\begin{acks}
We want to thank the participants of our formative and user studies for their time and feedback, as well as the anonymous reviewers for their insightful comments and constructive suggestions. We would also like to thank Jennifer Jacobs for the valuable discussions on the user evaluation design and analysis and Yunhao Luo for helping with building the user interface. 
\end{acks}

\bibliographystyle{ACM-Reference-Format}
\bibliography{dis2024}

\appendix
\section{Formative Study Interview Questions} \label{sec:formative_study_interview}
Kick-off:
\begin{itemize}
    \item How long have you been doing choreography? What genre(s)? 
\end{itemize}

\noindent Choreography:
\begin{itemize}
    \item What steps do you usually take to choreograph a dance sequence? Are there any tools you use?
    \begin{itemize}
        \item Follow-up: What difficulties, if any, have you encountered in the choreography process, and how do you address them? 
    \end{itemize}
    \item How do you stay creative for the choreography process? 
    \begin{itemize}
        \item Follow-up: What other things do you think could help you stay creative or enhance creativity? 
    \end{itemize}
    \item Do you ever choreograph dances in genres you are less familiar with? If yes, how do you prepare for that and go about doing it?
    \begin{itemize}
        \item Follow-up: What difficulties do you think you may encounter? What can be helpful to help you address the challenges? 
    \end{itemize}
    \item How do you evaluate a created dance sequence? Are there specific elements you look for in a choreographed sequence? Are there specific principles or guidelines to evaluate choreography, and if yes, what might those be? 
\end{itemize}

\noindent Reflection:
\begin{itemize}
    \item If there was a tool that could help with choreographing dance sequences,
    \begin{itemize}
        \item Would you want to use it? What would you like it to do?
        \item How would you want to interact with it? E.g., verbally, through text descriptions, sketching, and video. 
        \item In what aspects do you want it to help you? What output do you want to get? 
    \end{itemize}
    \item Is there anything else we have not covered that you would like to share?
\end{itemize}

\begin{figure*}[!ht]
    \centering
    \includegraphics[width=0.65\textwidth]{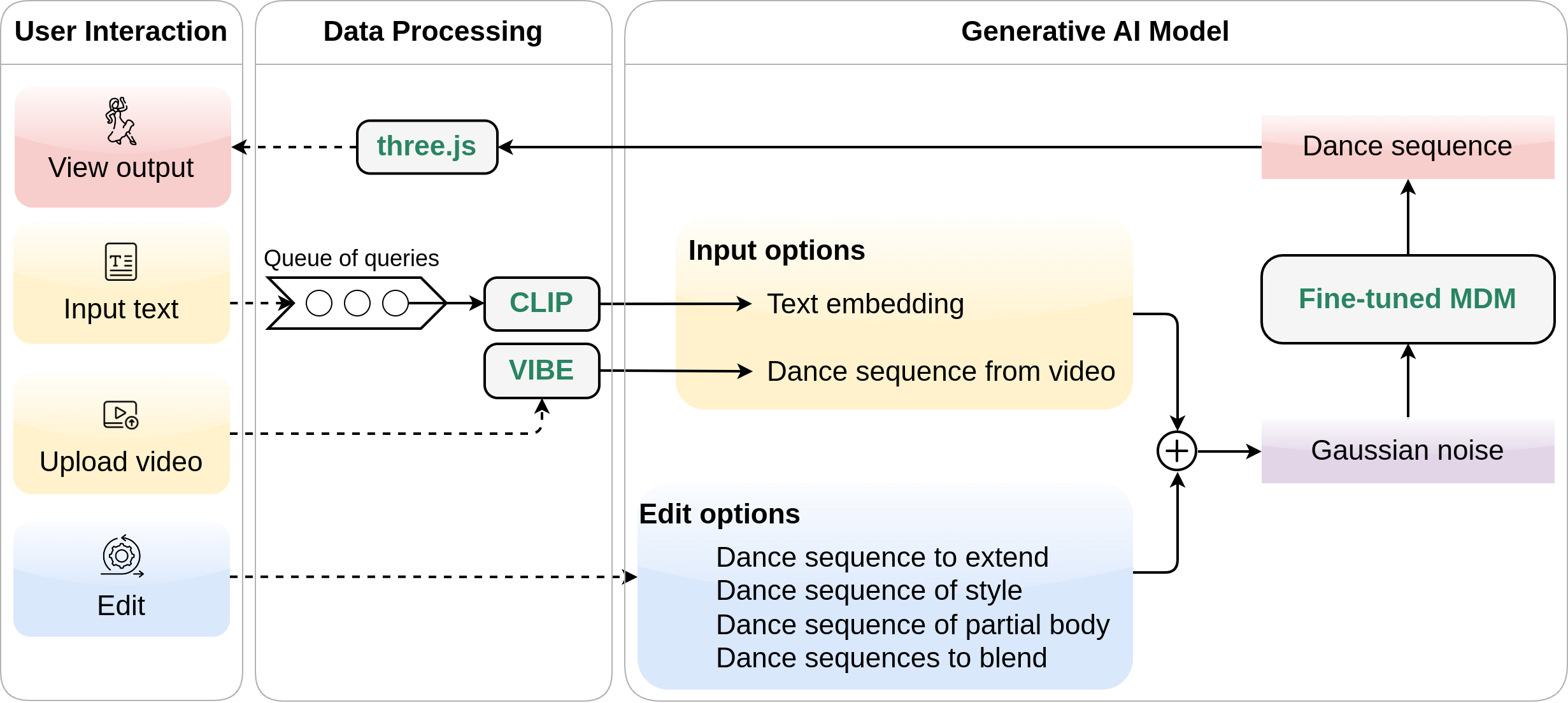}
    \caption{System technical architecture. Users interact with the user interface to provide text descriptions, video uploads, or direct edits. The user interface then transmits this information to the backend server for processing. Specifically, text descriptions are encoded as CLIP embeddings~\cite{radford2021learning} and uploaded videos are transformed into 3D sequences using VIBE~\cite{kocabas2020vibe}. All user input and edits are combined and fed into the fine-tuned MDM~\cite{tevet2023human} to generate dance sequences. The generated sequences are visualized within the user interface using three.js~\cite{threejs} to apply a mesh onto the dance motion. The system's user interface and backend server are connected via a TCP socket, as shown in the dashed arrows, while the dataflows within the backend server are represented as solid arrows.}
    \Description{This figure consists of the user interaction, data processing, and the backend AI model. The view of generated dance sequences, along with three user input types, is depicted within the user interaction. In the data processing column, the input text is encoded as CLIP text embeddings, and uploaded videos are converted to 3D using VIBE. In the generative AI model, the input and edit options are summed with a Gaussian noise sequence before input into the fine-tuned MDM to produce a sequence of dance movements.}
    \label{fig:technical_details}
\end{figure*}

\section{Technical Details} \label{sec:technical_details}
Figure~\ref{fig:technical_details} provides an overview of the technical architecture of the system that consists of user interaction, data processing, and a generative AI model. 

\subsection{User Interaction and Data Processing}
The user interface is built with HTML/CSS and JavaScript, allowing users to interact with the system. Users can either provide text descriptions of desired dance moves or upload video references.

\subsubsection{Text Input.}
Text descriptions are converted into queries, placed in a queue, and sent to the backend server via a TCP socket~\cite{tcp_server}. The backend leverages CLIP~\cite{radford2021learning} to encode these queries into text embeddings, which are then fed into the generative model to create a new dance sequence.

\subsubsection{Video Input.}
Uploading a dance video triggers the backend to use VIBE~\cite{kocabas2020vibe} to track the dancer and extract a 3D representation of the movement.

\subsubsection{Dance Sequence Management and Visualization.}
Each generated dance sequence is assigned a unique identifier. This ID is used for editing purposes. When a user selects a sequence for editing, the frontend sends the editing instructions to the backend. The backend retrieves the corresponding sequence based on its ID and feeds it into the generative model for modifications.

The final dance sequences, whether generated from text or edited by users, are returned to the user interface for display and interaction. We use the three.js library~\cite{threejs} to visualize these sequences within the web browser. The avatars showcasing the dance moves are SMPL meshes~\cite{SMPL_2015} (both male and female forms) or a Mixamo mesh~\cite{mixamo}.

\begin{figure*}[!ht]
    \centering
    \begin{subfigure}[b]{0.25\textwidth}
        \includegraphics[width=\textwidth]{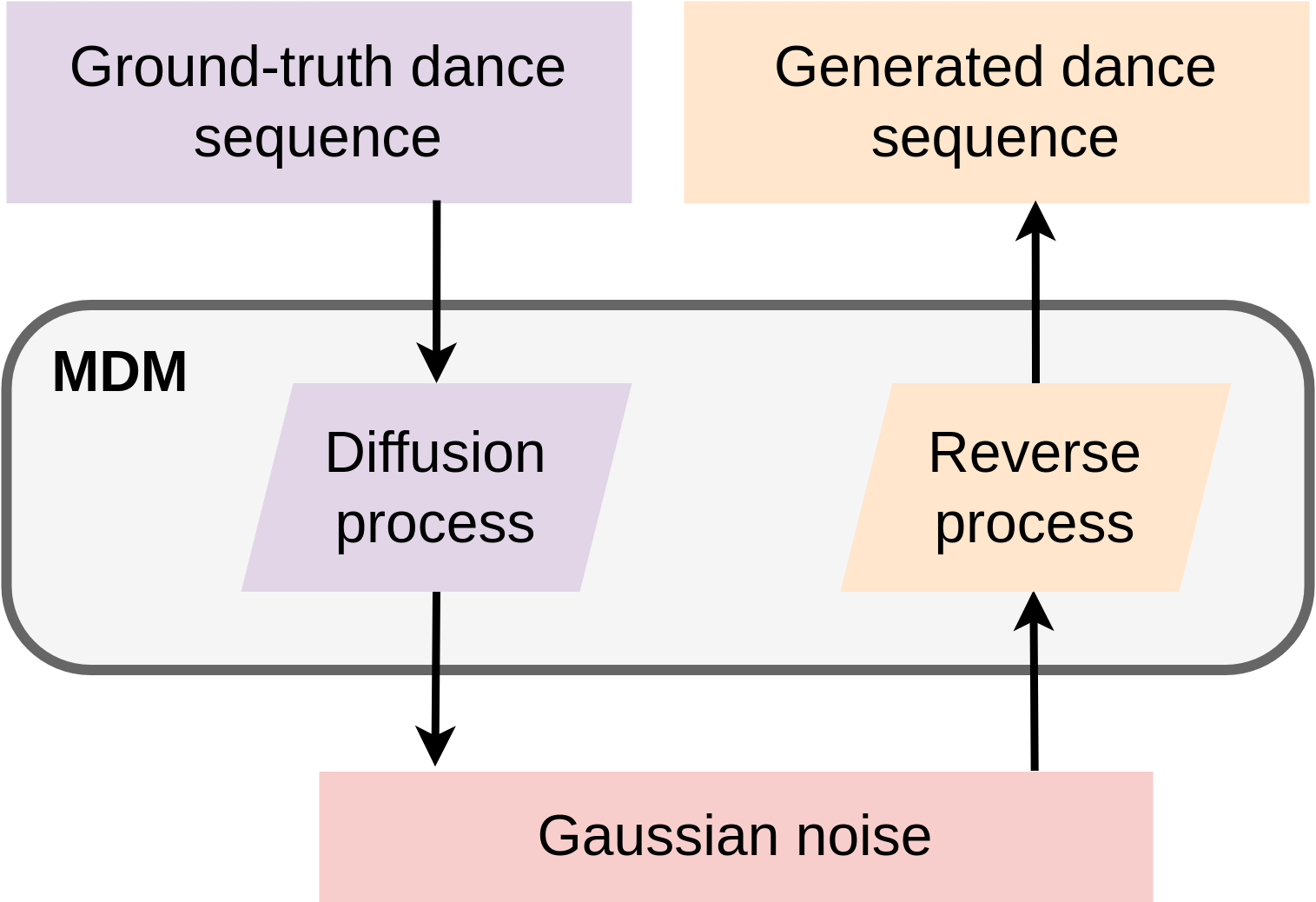}
        \caption{}
        \label{fig:algorithm_finetuning}
    \end{subfigure}
    \hspace{0.05\textwidth}
    \begin{subfigure}[b]{0.4\textwidth}
        \includegraphics[width=\textwidth]{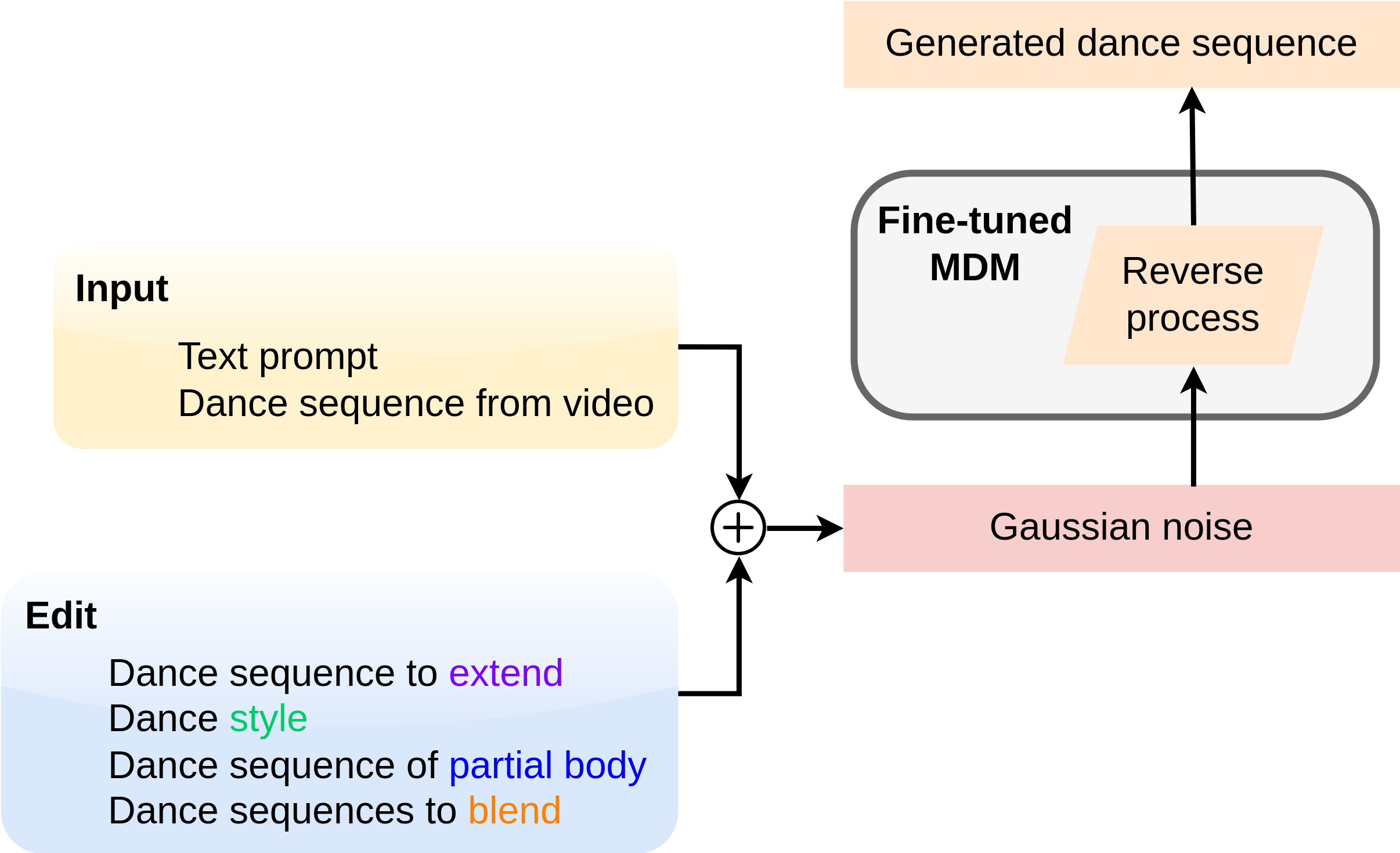}
        \caption{}
        \label{fig:algorithm_inference}
    \end{subfigure}
    \caption{The process of fine-tuning and inference for the dance sequence generation model. 
    During the fine-tuning stage (a), the pre-trained model, MDM~\cite{tevet2023human}, is fine-tuned using the AIST++ dataset~\cite{Li_2021_ICCV}. The ground-truth dance sequence from AIST++ undergoes a diffusion process, transforming it into a Gaussian noise sequence. This noise is then input into the reverse process, where denoising occurs, ultimately generating a dance sequence. 
    In the subsequent inference phase (b), the fine-tuned model processes user inputs for dance generation and editing. Text prompts are encoded as text embeddings using CLIP~\cite{radford2021learning}. Simultaneously, user-uploaded dance videos are transformed into 3D dance sequences using VIBE~\cite{kocabas2020vibe}. The text features or 3D dance sequences are combined with editing conditions for further refinement, encompassing dance sequences for expansion and blending, options for style changing, and partial body parts for editing. These inputs and conditions, along with a Gaussian noise sequence, are fed into the reverse process to generate and edit dance sequences. This comprehensive approach allows for user-controlled dance sequence creation and refinement.}
    \Description{There are two flow charts illustrating the fine-tuning (left) and inference (right) of the backend AI model. The fine-tuning figure displays a diffusion process that takes a ground-truth dance sequence from AIST++ as input and produces a Gaussian noise sequence, and a reverse process that takes the Gaussian noise sequence as input and generates a dance sequence. The inference figure demonstrates the input information and editing conditions are summed with a Gaussian noise sequence and fed into the fine-tuned model for dance sequence generation and editing. All these data flows are shown as solid arrows.}
    \label{fig:algorithm}
\end{figure*}

\subsection{Generative AI Model}
\subsubsection{Dance Sequence Generation.}
Our system utilizes the Motion Diffusion Model (MDM)~\cite{tevet2023human} for dance creation and editing. MDM was pre-trained on a massive dataset (HumanML3D~\cite{guo2022generating}) encompassing diverse everyday human movements, along with natural language descriptions. To further specialize in dance, we fine-tuned MDM using the AIST++ dance dataset~\cite{Li_2021_ICCV}. This dataset includes human dance motions categorized into ten distinct genres.

Figure~\ref{fig:algorithm} illustrates the model fine-tuning and inference processes. During fine-tuning, we froze the CLIP model for text input encoding and trained the remaining layers of the model using the training set of AIST++. The learning rate was 5e-5, the batch size was 16, and the number of epochs was 50. The rest of the training parameters remained the same as the pre-trained MDM. The fine-tuning took roughly 13 hours on a single NVIDIA RTX 3090 GPU. 

\subsubsection{Dance Sequence Style Control.}
While our underlying AI model lacks inherent style representation, we aimed to enable motion style transfer. Previous work by SinMDM~\cite{raab2023single} introduced similar styles within a motion harmonization context. However, directly describing styles in text prompts, such as emotions, would not effectively influence dance styles in our system due to the limited style labels in the training data. To overcome this limitation, we adopted a two-step approach: 
\begin{enumerate}
    \item Style definition: We explicitly defined six distinct styles based on the characteristics present within the HumanML3D dataset.
    \item Style transfer models: Leveraging a training method similar to SinMDM, we trained separate models to obtain reference sequences for each style, enabling the system to translate styles into dance sequences for style control. This process involves extracting the high-frequency information of the source dance sequence and adding it to the reference sequence. Specifically, the reference sequence $y$ undergoes a low-pass filter $\phi$ (proposed by~\cite{choi2021ilvr}) to obtain its low-frequency features $\phi(y)$. The high-frequency features of the source sequence $x$ are obtained by $x - \phi(x)$. The resulting sequence, which combines the high-frequency details from the source sequence and the low-frequency information describing style from the reference sequence, is obtained by $x - \phi(x) + \phi(y)$. 
\end{enumerate}

\section{User Study Interview Questions} \label{sec:user_study_interview}
Background:
\begin{itemize}
    \item What dance genre(s) do you choreograph? How long have you been choreographed?
\end{itemize}

\noindent System features:
\begin{enumerate}
    \item Input modality:
    \begin{itemize}
        \item Was it helpful for you to describe what you want to create in natural language/text as input? Why or why not?
        \item Was it helpful for you to use video as input to generate choreography? Why or why not?
        
        Follow-up: 
        \begin{itemize}
            \item Was there a particular reason you started with text or with video? What is the usual starting point in your choreography creation process after you have an idea in mind?
            \item What types of mediums do you currently use to communicate your ideas and early choreography sequences? How often do you use them?
            \item What would you like to be able to do when it comes to sharing your early ideas with others for feedback and collaborative brainstorming?
        \end{itemize}
    \end{itemize}

    \item Ideation:
    \begin{itemize}
        \item Our system outputs three results each time you make a request for a dance sequence. Did you find that to be a helpful way to explore dance ideas? Could you please expand on why or why not?
        \item How important is it to you to be able to see multiple variations of a generated choreography based on your text prompt or video input? Would being able to see multiple variations of the generated output be helpful in your choreography creation process?
        \item Did you get any results that made you feel interesting? If yes, what do you think about these results? 
    \end{itemize}

    \item Iterative prototyping:
    \begin{itemize}
        \item How important is it to you to be able to iterate in your choreography?
        \item Do you find editing dance sequences helpful for you to iterate?
    \end{itemize}

    \item Documentation:
    \begin{itemize}
        \item What do you think about the documentation methods? Did you find them helpful when you interact with the system? Why?
        \item Did you download any dance sequences you created? If yes, how may you use the results you exported from the system?
    \end{itemize}

    \item Creative process:
    \begin{itemize}
        \item Are there specific tasks in your current process that this system could help with? If yes, in what way(s) would it be helpful?

        Follow-up: 
        \begin{itemize}
            \item How important is such a process in your choreography?
            \item Why do you want such a process to be supported?
        \end{itemize}

        \item What scenarios do you think this system could fit in your workflow? 
    \end{itemize}
\end{enumerate}

\noindent Others:
\begin{itemize}
    \item How do you want this system to be personalized for your creative work/workflow?
    \item How much control do you think you have over the outcomes? Do you think your current control is enough/appropriate? Do you think some automatic is helpful/bad?
    \item Was there something you could not do or had difficulty doing before but can now do using this system? If so, could you share some details about it?
    \item What aspects do you think this system could be improved to fit your creative process better?
    \item Is there anything else you’d like us to know that we haven’t talked about?
\end{itemize}

\end{document}